\documentclass[a4paper,10pt]{article}
\usepackage{graphicx}
\usepackage{comment}
\usepackage{amssymb}
\usepackage{amsmath}
\usepackage{nicefrac}
\usepackage{graphicx}
\usepackage{dcolumn}
\usepackage{bm}
\usepackage{comment}
\usepackage{multirow}
\usepackage{cite}
\usepackage{mathrsfs}
\usepackage{url}

\newcommand{\ab}{\alpha\rightarrow\beta}
\newcommand{\ga}{g_\alpha}
\newcommand{\gb}{g_\beta}
\newcommand{\vgab}{\mathcal{V}_G^{\ab}(\vec{p})}
\newcommand{\vg}{\mathcal{V}_G}

\newcommand{\vgaab}{\mathcal{V}_G^{\alpha,\ab}}
\newcommand{\vgbab}{\mathcal{V}_G^{\beta,\ab}}
\newcommand{\vgrab}{\mathcal{V}_G^{r,\ab}}
\newcommand{\vgsab}{\mathcal{V}_G^{s,\ab}}
\newcommand{\bin}[2]{\left(\begin{array}{c} \!\!\!#1\!\!\! \\  \!\!\!#2\!\!\! \end{array}\right)}
\newcommand{\DO}{D_0^{\alpha\beta}}
\newcommand{\Dr}{D_r^{\alpha\beta}}

\newcommand{\Dt}{D_t^{\alpha\beta}}
\newcommand{\threej}[6]{\left(\begin{array}{ccc}#1 & #2 & #3 \\ #4 & #5 & #6 \end{array}\right)}
\newcommand{\sixj}[6]{\left\{\begin{array}{ccc}#1 & #2 & #3 \\ #4 & #5 & #6 \end{array}\right\}}
\newcommand{\neufj}[9]{\left\{\begin{array}{ccc}#1 & #2 & #3 \\ #4 & #5 & #6 \\ #7 & #8 & #9 \end{array}\right\}}

\def\anglefig{0}
\def\scalefig{0.90}

\begin{document}

\huge

\begin{center}
On the computation of moments in the Super-Transition-Arrays model for radiative opacity calculations
\end{center}

\vspace{0.3cm}

\large

\begin{center}
Jean-Christophe Pain$^{a,b,}$\footnote{jean-christophe.pain@cea.fr}
\end{center}

\normalsize

\begin{center}
\it $^a$CEA, DAM, DIF, F-91297 Arpajon, France\\
\it $^b$Universit\'e Paris-Saclay, CEA, Laboratoire Mati\`ere en Conditions Extr\^emes,\\
\it 91680 Bruy\`eres-le-Ch\^atel, France
\end{center}

\vspace{0.3cm}

\large

\begin{center}
Brian Wilson\footnote{wilson9@llnl.gov}
\end{center}

\normalsize

\begin{center}
\it Lawrence Livermore National Laboratory, P.O. Box 808, L-414, Livermore, California 94551, USA
\end{center}

\vspace{0.3cm}

\begin{abstract}
In the Super-Transition-Array statistical method for the computation of radiative opacity of hot dense matter, the moments of the absorption or emission features involve partition functions with reduced degeneracies, occurring through the calculation of averages of products of subshell populations. In the present work, we discuss several aspects of the computation of such peculiar partition functions, insisting on the precautions that must be taken in order to avoid numerical difficulties. In a previous work, we derived a formula for supershell partition functions, which takes the form of a functional of the distribution of energies within the supershell and allows for fast and accurate computations, truncating the number of terms in the expansion. The latter involves coefficients for which we obtained a recursion relation and an explicit formula. We show that such an expansion can be combined with the recurrence relation for shifted partition functions. We also propose, neglecting the effect of fine structure as a first step, a positive-definite formula for the Super-Transition-Array moments of any order, providing an insight into the asymmetry and sharpness of the latter. The corresponding formulas are free of alternating sums. Several ways to speed up the calculations are also presented.
\end{abstract}

\section{Introduction}\label{sec1}

The radiative opacity of warm/hot plasmas is of primary importance for designing and analyzing inertial confinement fusion (ICF) experiments \cite{Goldstein2012,Kraus2016} as well as for understanding astrophysical observations and phenomena \cite{Clayton1968,Sturrock1986,Barnes2013,Cowperthwaite2017}. The radiation-transport properties of these plasmas, among which the absorption spectrum, are determined, mainly, by the electronic structure of the ions together with the photon distribution. Experimental opacity measurements are rather difficult to carry out, for plasmas in uniform density and temperature conditions \cite{Kurzweil2022}. Furthermore, ICF experiments and astrophysical phenomena require opacity in a very wide density–temperature regime, for various materials. Therefore, the opacity is traditionally determined by theoretical models. The latter have to tackle a highly complicated physical problem of coupled atom–plasma systems, in order to calculate the absorption spectrum. Also, the models should have the ability to sum up the whole relevant electronic quantum transitions, from all of the bound and free electronic configurations of the plasma ions. This very hard problem requires approximations, and each opacity numerical code has its own assumptions and numerical methods. Bar-shalom \emph{et al.} developed a powerful technique for the calculation of bound-bound and bound-free photo-absorption spectra \cite{Barshalom1989}. It consists in grouping the huge number of lines between the enormous number of configurations into large sets, called Super-Transition-Arrays (STAs). The STA moments are calculated analytically via the partition function algebra, and are split into smaller STAs until convergence is achieved. This procedure enables the handling of situations where the number of populated configurations is too large \cite{Pain2020a}. Krief discussed two useful methods for the estimation of the number of populated configurations: using an exact calculation of the total combinatoric number of configurations within superconfigurations in a converged STA calculation on one hand, and by using an estimate for the multidimensional width of the probability distribution for electronic population over bound shells (which is binomial if electron exchange and correlation effects are neglected) on the other hand \cite{Krief2021}. There is still a great interest in STA-based opacity codes, as evidenced by the recent publications on the subject (see for instance Refs. \cite{Hazak2013,Kurzweil2013,Wilson2015,Pain2015,Kurzweil2016,Krief2016,Krief2018,Pain2021,Aberg2021,Pain2021,Pain2022,Gill2023,Gill2023b}).

The partition function of a supershell with $N$ subshells and $Q$ electrons is defined as
\begin{equation}
\mathcal{U}_Q(g)=\sum_{\substack{\{p_s\}\\\sum_{s=1}^Np_s=Q}}\prod_{s=1}^N\bin{g_s}{p_s}X_s^{p_s},
\end{equation}
where $X_s=e^{-\beta(\epsilon_s-\mu)}$, with the usual notation $\beta=1/(k_BT)$, $k_B$ being the Boltzmann constant. $\epsilon_s$ is the energy of subshell $s$, $p_s$ its population and $\mu$ the chemical potential. We use the notation
\begin{equation}
\sum_{\substack{\{p_s\}\\\sum_{s=1}^Np_s=Q}}\cdots=\underbrace{\sum_{p_1=0}^{g_1}\sum_{p_2=0}^{g_2}\cdots \sum_{p_N=0}^{g_N}}_{\sum_{s=1}^Np_s=Q}\cdots.
\end{equation}
Partition functions with reduced degeneracies are important for the calculation of the STA moments. By reduced degeneracies, we mean that the degeneracy of one or more subshells in the supershell is reduced by one or more. The reduced degeneracies are sometimes called ``shifted statistical weights'' \cite{Oreg1997} or ``modified'' degeneracies \cite{Barshalom1997}. They are noted in the following way: $\mathcal{U}_Q\left(g^a\right)$ means that the set of degeneracies $\{g_1,g_2,\cdots,g_N\}$ is replaced by $\{g_1,g_2,\cdots g_a-1,\cdots,g_N\}$ in the computation of the partition function. In the same way, $\mathcal{U}_Q\left(g^{abbc}\right)$ means that the set of degeneracies $\{g_1,g_2,\cdots,g_N\}$ is replaced by $\{g_1,g_2,\cdots g_a-1,\cdots,g_b-2,\cdots,g_c-1,\cdots,g_N\}$.

In section \ref{sec2}, we discuss the importance of partition functions with shifted degeneracies in the calculation of the STA moments, insisting on the different ways of computing them in order to circumvent numerical difficulties. We insist on the precautions that must be taken, especially for non-relativistic ``s'' ($\ell=0$) or relativistic p$_{1/2}$ ($j=1/2$) subshells (for which a twice-reduced degeneracy falls to zero), and address the long-running issue of alternating sums.

In a previous work, we derived a formula for supershell partition functions, which takes the form of a functional of the distribution of energies within the supershell and allows for fast and accurate computations, truncating the number of terms in the expansion. The latter involves coefficients (denoted $\Gamma_k$) for which we obtained a recursion relation and an explicit formula. In section \ref{sec3}, we show that such a relation applies also for partition functions with shifted degeneracies, and provide the corresponding expressions.

Finally, in section \ref{sec4}, we propose, discarding the contribution of fine-structure, an expression for the $n^{th}$-order STA moment avoiding alternating sums. The formula involves multinomial coefficients and Stirling numbers of the second kind. The numerical efficiency of such an evaluation is connected to the enumeration of partitions. In the same vein, we mention several numerical optimizations such as fast exponentiation and Karatsuba-type algorithms for the product of multivariate polynomials, which should be of great interest for the computation of opacity in the superconfiguation approximation. A matrix representation of the recurrence relations, acting on vectors of partition functions, is described.

\section{Reduced degeneracies in the STA theory}\label{sec2}

\subsection{General relations and mean populations}\label{subsec21}

The mean population of subshell $a$ inside a $Q$-electron subshell is
\begin{equation}
\langle p_a\rangle=\frac{1}{\mathcal{U}_Q(g)}\sum_{\substack{\{p_s\}\\\sum_{s=1}^Np_s=Q}}p_a\prod_{s=1}^N\bin{g_s}{p_s}X_s^{p_s}.
\end{equation}
Using the identity
\begin{equation}
p\bin{g}{p}=g\bin{g-1}{p-1},
\end{equation}
the average population of subshell $a$ reads
\begin{equation}
\langle p_a\rangle=\frac{1}{\mathcal{U}_Q(g)}\sum_{\substack{\{p_s\}\\\sum_{s=1}^Np_s=Q}}g_a\prod_{s=1}^N\bin{g_s-\delta_{s,a}}{p_s-\delta_{s,a}}X_s^{p_s}.
\end{equation}
Then, using
\begin{equation}\label{bin2}
\bin{g}{p}=\bin{g-1}{p}+\bin{g-1}{p-1},
\end{equation}
we get
\begin{equation}
\langle p_a\rangle=\frac{g_a}{\mathcal{U}_Q(g)}\sum_{\substack{\{p_s\}\\\sum_{s=1}^Np_s=Q}}\prod_{s=1}^N\bin{g_s}{p_s}X_s^{p_s}-\frac{g_a}{\mathcal{U}_Q(g)}\sum_{\substack{\{p_s\}\\\sum_{s=1}^Np_s=Q}}\prod_{s=1}^N\bin{g_s-\delta_{s,a}}{p_s}X_s^{p_s},
\end{equation}
\emph{i.e.}
\begin{equation}
\langle p_a\rangle=\frac{g_a}{\mathcal{U}_Q(g)}\mathcal{U}_Q(g)-\frac{g_a}{\mathcal{U}_Q(g)}\mathcal{U}_Q\left(g^a\right),
\end{equation}
yielding
\begin{equation}\label{pop1}
\langle p_a\rangle=g_a\left[1-\frac{\mathcal{U}_Q\left(g^a\right)}{\mathcal{U}_Q(g)}\right].
\end{equation}

Applying the relation (\ref{bin2}) again, one gets
\begin{equation}
\mathcal{U}_Q(g)=\sum_{\substack{\{p_s\}\\\sum_{s=1}^Np_s=Q}}\prod_{s=1}^N\bin{g_s}{p_s}X_s^{p_s}=\sum_{\substack{\{p_s\}\\\sum_{s=1}^Np_s=Q}}\prod_{s=1}^N\bin{g_s-\delta_{s,a}}{p_s}X_s^{p_s}+\sum_{\substack{\{p_s\}\\\sum_{s=1}^Np_s=Q-1}}\prod_{s=1}^N\bin{g_s-\delta_{s,a}}{p_s-\delta_{s,a}}X_s^{p_s}
\end{equation}
or
\begin{equation}
\mathcal{U}_Q(g)=\mathcal{U}_Q\left(g^a\right)+X_a\sum_{\substack{\{p_s\}\\\sum_{s=1}^Np_s=Q-1}}\prod_{s=1}^N\bin{g_s-\delta_{s,a}}{p_s}X_s^{p_s}
\end{equation}
yielding the relation
\begin{equation}\label{rel1}
\mathcal{U}_Q(g)=\mathcal{U}_Q\left(g^a\right)+X_a\mathcal{U}_{Q-1}\left(g^a\right),
\end{equation}
from which it follows that
\begin{equation}\label{troitroi}
\mathcal{U}_Q\left(g^a\right)=\sum_{k=0}^Q\mathcal{U}_{Q-k}(g)(-X_a)^k.
\end{equation}
Applying Eq. (\ref{troitroi}) for $\mathcal{U}_{Q-k}\left(g^b\right)$, we obtain
\begin{equation}
\mathcal{U}_{Q-k}\left(g^b\right)=\sum_{r=0}^{Q-k}\mathcal{U}_{Q-k-r}(g)(-X_b)^r,
\end{equation}
yielding
\begin{equation}
\mathcal{U}_Q\left(g^{ab}\right)=\sum_{k=0}^Q\sum_{r=0}^{Q-k}(-X_a)^k(-X_b)^r\mathcal{U}_{Q-k-r}(g).
\end{equation} 
Setting $m=k+r$, we get
\begin{equation}
\mathcal{U}_Q\left(g^{ab}\right)=\sum_{k=0}^Q\sum_{m=k}^Q(-X_a)^k(-X_b)^{m-k}\mathcal{U}_{Q-m}(g),
\end{equation} 
\emph{i.e.},
\begin{eqnarray}\label{troiquoi}
\mathcal{U}_Q\left(g^{ab}\right)&=&\sum_{k=0}^Q\left[\sum_{m=0}^k(-X_a)^m(-X_b)^{k-m}\right]\mathcal{U}_{Q-k}(g)\nonumber\\
&=&\sum_{k=0}^Q(-X_b)^k\left[\sum_{m=0}^k\left(\frac{X_a}{X_b}\right)^m\right]\mathcal{U}_{Q-k}(g),
\end{eqnarray} 
giving finally
\begin{equation}\label{troicinq}
\mathcal{U}_Q\left(g^{ab}\right)=\sum_{k=0}^Q(-1)^k\left[\frac{X_b^{k+1}-X_a^{k+1}}{X_b-X_a}\right]\mathcal{U}_{Q-k}(g).
\end{equation} 

Using Eq. (\ref{rel1}), we can write
\begin{equation}\label{pop2}
\langle p_a\rangle=g_aX_a\frac{\mathcal{U}_{Q-1}\left(g^a\right)}{\mathcal{U}_Q(g)},
\end{equation}
or also
\begin{equation}\label{pop3}
\langle p_a\rangle=\frac{g_a}{1+\displaystyle\frac{\mathcal{U}_Q\left(g^a\right)}{X_a\mathcal{U}_{Q-1}\left(g^a\right)}}.
\end{equation}
Expressions (\ref{pop1}), (\ref{pop2}) and (\ref{pop3}) are equivalent. 

In order to reduce the computation time, it is interesting, when the number of electrons in a supershell is larger than half of its degeneracy, to deal with the ``complementary'' of electrons: the ``holes''. The definition of a partition function in terms of holes, noted $\mathcal{U}^*$, is obtained from the partition function in terms of electrons $\mathcal{U}$, under the substitution $X_k\rightarrow X^*_k=1/X_k$. The same recurrence relations applies. The number of electrons $Q$ is replaced by the number of holes $Q^*=G-Q$, where $G$ is the degeneracy of the supershell defined by
\begin{equation}
G=\sum_{s=1}^Ng_s. 
\end{equation}
Relation (\ref{rel1}) becomes
\begin{equation}\label{rel1t}
\mathcal{U}^*_{Q^*}(g)=X^*_a\mathcal{U}^*_{Q^*-1}\left(g^a\right)+\mathcal{U}^*_{Q^*}\left(g^a\right).
\end{equation}

As concerns reduced degeneracies, we have the following relations
\begin{equation}
\mathcal{U}_{Q-n}(g)=\mathcal{A}~\mathcal{U}^*_{Q^*+n}(g),
\end{equation}
as well as
\begin{equation}
\mathcal{U}_{Q-n}(g)=\mathcal{A}~X^*_a\mathcal{U}^*_{Q^*+n-1}\left(g^a\right),
\end{equation}
and
\begin{equation}
\mathcal{U}_{Q-n}\left(g^{ab}\right)=\mathcal{A}~X^*_aX^*_b\mathcal{U}^*_{Q^*+n-2}\left(g^{ab}\right),
\end{equation}
with
\begin{equation}
\mathcal{A}=\prod_{s=1}^NX_s^{g_s}.
\end{equation}
Within the ``hole'' formalism, one has therefore
\begin{equation}\label{pop2t}
\langle p_a\rangle=\langle g_a-p_a\rangle^*=g_a\frac{\mathcal{U}^*_{Q^*}\left(g^a\right)}{\mathcal{U}^*_{Q^*}(g)},
\end{equation}
or also
\begin{equation}\label{pop3t}
\langle p_a\rangle=\frac{g_a}{1+X^*_a\displaystyle\frac{\mathcal{U}^*_{Q^*-1}\left(g^a\right)}{\mathcal{U}^*_{Q^*}\left(g^a\right)}}.
\end{equation}
For holes, we iterate from
\begin{equation}
\mathcal{U}^*_{Q^*}\left(g^a\right)=\frac{1}{X_a}\left[\mathcal{U}^*_{Q^*+1}(g)-\mathcal{U}^*_{Q^*+1}\left(g^a\right)\right],
\end{equation}
yielding
\begin{equation}\label{troiquat}
\mathcal{U}^*_{Q^*}\left(g^a\right)=-\sum_{n=1}^{G-Q^*}\mathcal{U}^*_{Q^*+n}(g)(-X_a)^{-n},
\end{equation}
where $G=\sum_{s=1}^Ng_s$. Applying Eq. (\ref{troiquat}) twice, we get \cite{Krief2015}:
\begin{equation}
\mathcal{U}^*_{Q^*}\left(g^{ab}\right)=\sum_{k=2}^{G-Q^*}\left[\sum_{m=1}^{k-1}(-X_b)^{-(k-m)}(-X_a)^{-m}\right]\mathcal{U}^*_{Q^*+k}(g),
\end{equation}
which is equal to
\begin{equation}\label{troisix}
\mathcal{U}^*_{Q^*}\left(g^{ab}\right)=\sum_{k=2}^{G-Q^*}(-X_b)^{-k}X_a^{2-k}\left[\frac{X_a^{k-1}-X_b^{k-1}}{X_a-X_b}\right]\mathcal{U}^*_{Q^*+k}(g).
\end{equation}
Equations (\ref{troitroi}), (\ref{troicinq}), (\ref{troiquat}) and (\ref{troisix}) significantly simplify the calculations as they include usual partition function (\emph{i.e.}, with the entire set of degeneracies) $\mathcal{U}_k(g)$, whatever $a$ and $b$ \cite{Krief2015}. This is due to the fact that in the calculation of $\mathcal{U}_Q(g)$ (see section \ref{subsec31}), the series $\left\{\mathcal{U}_0(g), \cdots, \mathcal{U}_Q(g)\right\}$ or $\left\{\mathcal{U}_G(g), \cdots, \mathcal{U}_0(g)\right\}$ are calculated as well.

\begin{figure}
\begin{center}
\includegraphics[scale=0.5]{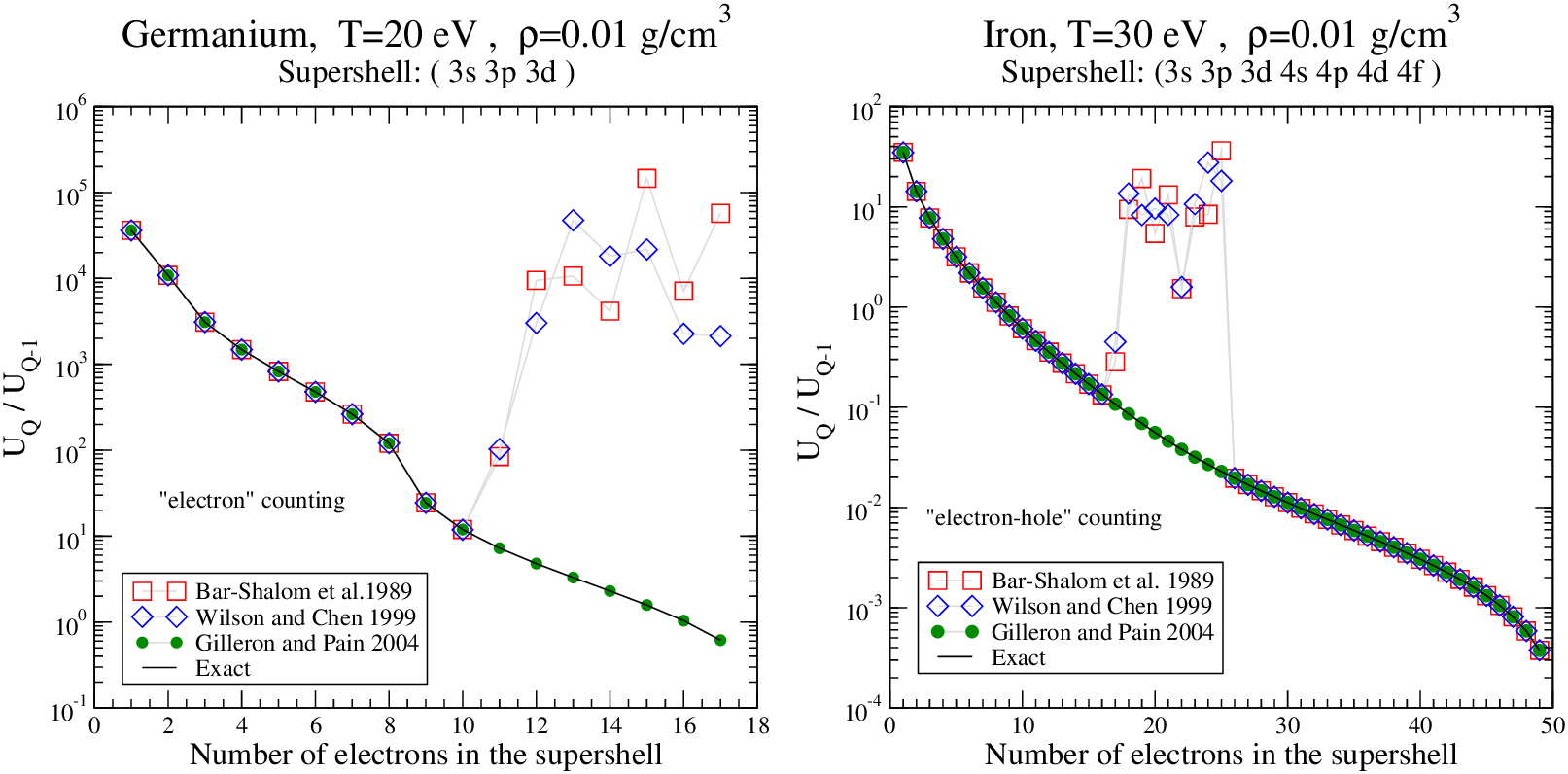}
\caption{Ratios of consecutive partition functions in two cases: supershell (3s3p3d) in a germanium plasma at $T=20$ eV and $\rho$=0.01 g/cm$^3$ (left), and supershell (3s3p3d4s4p4d4f) in an iron plasma at $T=30$ eV and $\rho$=0.01 g/cm$^3$ (right). Results of the recursion relations of Bar-Shalom \textit{et al.}, Wilson and Chen, and Gilleron and Pain (Eq. (\ref{gilpain})) compared to the exact values.}\label{brux2}
\end{center}
\end{figure}

\begin{figure}[!ht]
 \begin{minipage}[c]{0.48\textwidth}
   \centering
   \includegraphics[width=\textwidth,angle=\anglefig,scale=\scalefig]{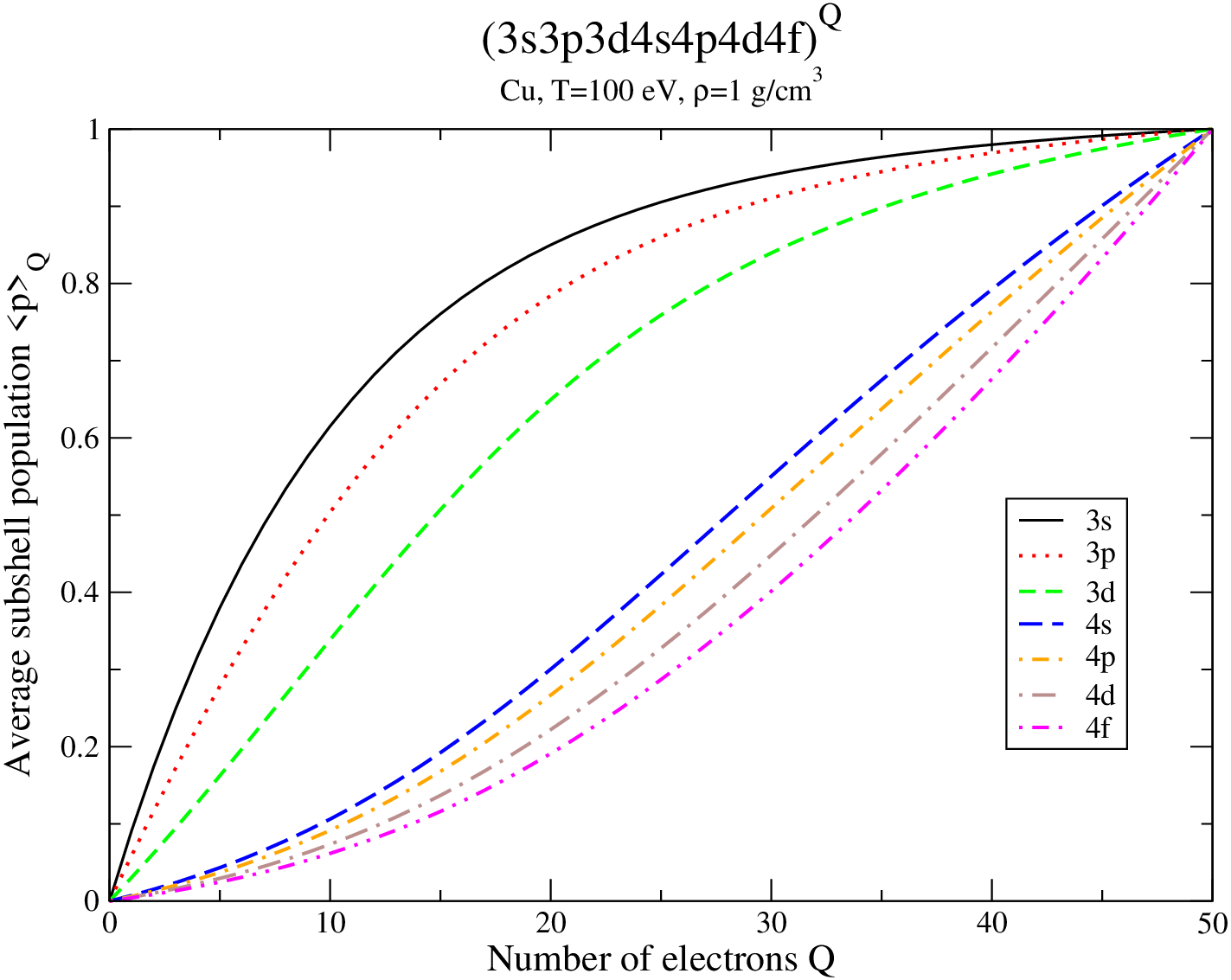}
   \caption{Average populations in supershell (3s3p3d) in a copper plasma at $T=$100 eV and $\rho=$1 g/cm$^3$. Results obtained with the recursion relation (\ref{gilpain}).}\label{Fig2}
 \end{minipage}\hfill
 \begin{minipage}[c]{0.48\textwidth}
   \centering
   \includegraphics[width=\textwidth,angle=\anglefig,scale=\scalefig]{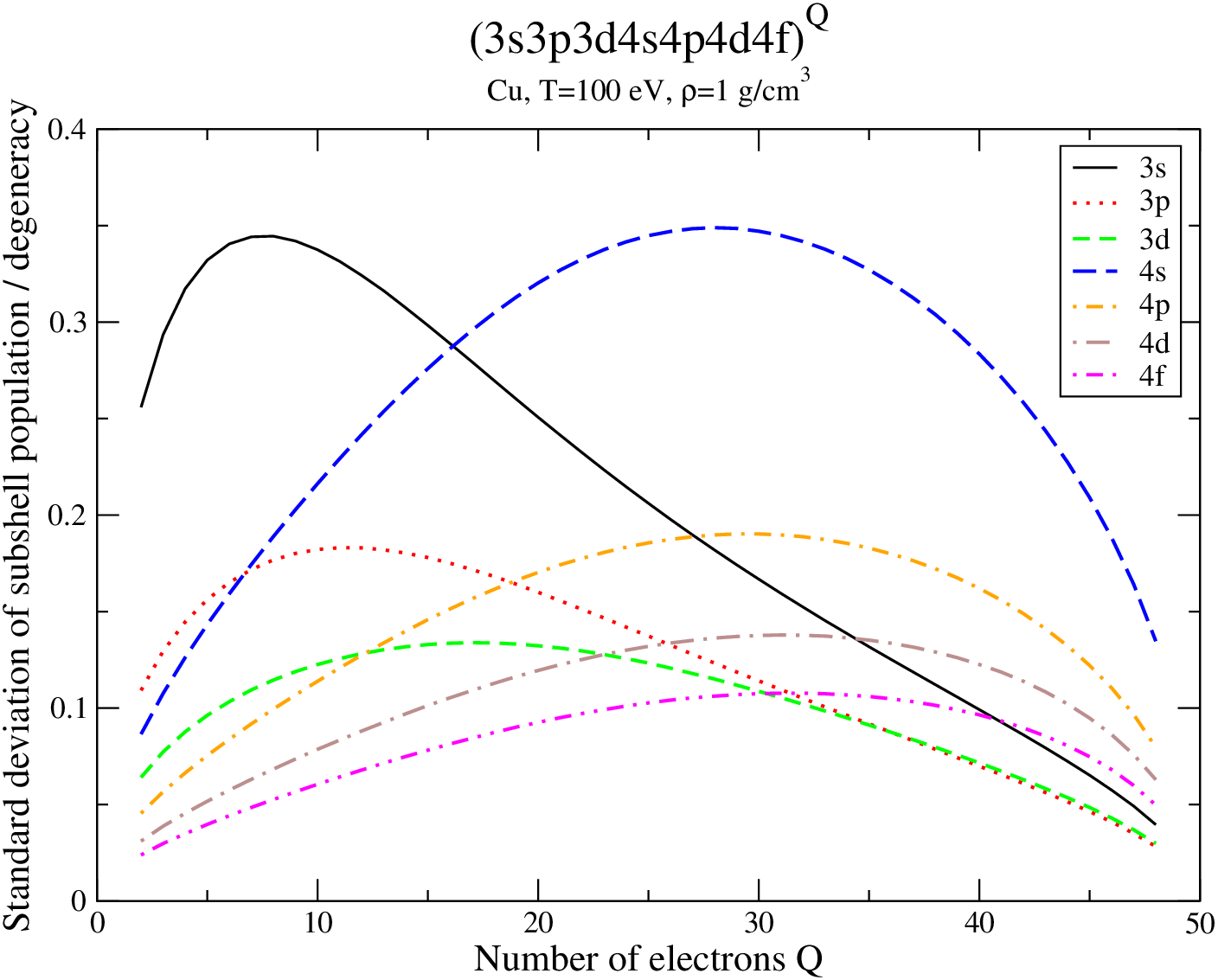}
   \caption{(Color online) Variances of populations in supershell (3s3p3d) in a copper plasma at $T=$100 eV and $\rho=$1 g/cm$^3$. Results obtained with the recursion relation (\ref{gilpain}).}\label{Fig3}
 \end{minipage}
\end{figure}

\begin{figure}
\begin{center}
\includegraphics[scale=0.4]{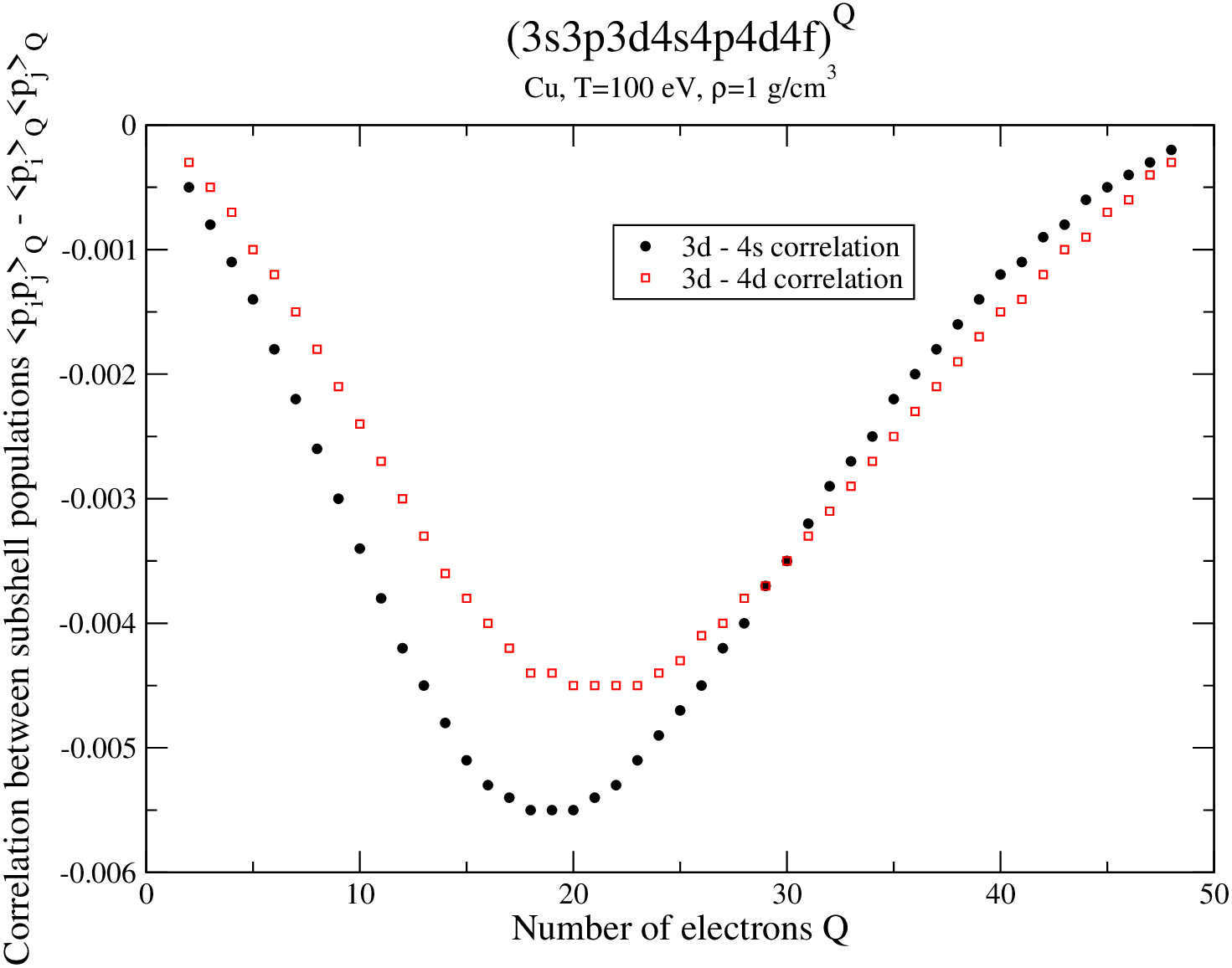}
\caption{Correlations between populations of orbitals 3d and 4s (full black circles) as well as between 3d and 4d (empty red squares) in supershell (3s3p3d) in a copper plasma at $T=$100 eV and $\rho=$1 g/cm$^3$. Results obtained with the recursion relation (\ref{gilpain}).}\label{Fig4}
\end{center}
\end{figure}

\subsection{Definitions}

Let us consider $N$ subshells, with degeneracies $g_1, g_2, \cdots, g_N$ populated by $Q$ electrons. We define
\begin{equation}
D_0^{\alpha\beta}=\langle\beta\rangle-\langle\alpha\rangle,
\end{equation}
where $\langle s\rangle$ is the one-electron integral 
\begin{equation}
\langle s\rangle=\epsilon_s+\langle s|-\frac{Z}{r}-V(r)|s\rangle,
\end{equation}
and
\begin{equation}
D_s^{\alpha\beta}=\langle\beta,s\rangle-\langle\alpha,s\rangle+\eta_{\alpha\beta}\left(\frac{\delta_{\alpha, s}}{g_{\alpha}-1}-\frac{\delta_{\beta, s}}{g_{\beta}-1}\right),
\end{equation}
where
\begin{equation}
\eta_{\alpha\beta}=\left(2\ell_{\alpha}+1\right)\left(2\ell_{\beta}+1\right)\left\{\sum_{k>0}f_kF^{(k)}\left(n_{\alpha}\ell_{\alpha},n_{\beta}\ell_{\beta}\right)+\sum_{k}g_kG^{(k)}\left(n_{\alpha}\ell_{\alpha},n_{\beta}\ell_{\beta}\right)\right\}
\end{equation}
with
\begin{equation}
f_k=\threej{\ell_{\alpha}}{k}{\ell_{\alpha}}{0}{0}{0}\threej{\ell_{\beta}}{k}{\ell_{\beta}}{0}{0}{0}\sixj{\ell_{\alpha}}{k}{\ell_{\alpha}}{\ell_{\beta}}{1}{\ell_{\beta}}
\end{equation}
and
\begin{equation}
g_k=\threej{\ell_{\alpha}}{k}{\ell_{\beta}}{0}{0}{0}^2\left\{\frac{2}{3}\delta_{k,1}-\frac{1}{2(2\ell_{\alpha}+1)(2\ell_{\beta}+1)}\right\},
\end{equation}
$F^{(k)}$ and $G^{(k)}$ being the direct and exchange Slater integrals respectively, and $\threej{\ell_1}{\ell_2}{\ell_3}{m_1}{m_2}{m_3}$ and $\sixj{\ell_1}{\ell_2}{\ell_3}{\ell_4}{\ell_5}{\ell_6}$ the Wigner $3j$ and Racah $6j$ symbols respectively. For $r\ne s$, the electron-electron interaction matrix elements are:
\begin{equation}
\langle r,s\rangle=F^{(0)}\left(n_r\ell_r,n_s\ell_s\right)-\frac{1}{2}\sum_k\threej{\ell_r}{k}{\ell_s}{0}{0}{0}^2G^{(k)}(n_r\ell_r,n_s\ell_s)
\end{equation}
and
\begin{equation}
\langle r,r\rangle=F^{(0)}\left(n_r\ell_r,n_r\ell_r\right)-\frac{(2\ell_r+1)}{(4\ell_r+1)}\sum_{k>0}\threej{\ell_r}{k}{\ell_r}{0}{0}{0}^2F^{(k)}(n_r\ell_r,n_r\ell_r).
\end{equation}
Note that $\langle s\rangle$ and $\langle r,s\rangle$ are sometimes noted $I_s$ and $V_{rs}$ respectively.

\subsection{Expression of the STA moments}\label{subsec22} 

For intermediate- or high-$Z$, the number of lines between two configurations $c$ and $c'$ can be so large that the average separation between two lines is smaller than the individual line widths. In this case, for a given single-electron jump, the lines coalesce into a quasi-Gaussian envelope called UTA (Unresolved Transition Array) \cite{Bauche1979}. 

In the STA theory \cite{Barshalom1989,Barshalom1991}, the unresolved transition arrays between two superconfigurations are also modeled by a continuous distribution of photon energy (often Gaussian), characterized by its moments of order 0, 1 and 2, resulting from an average of the inter-term transition elements over all the configurations involved. In the same way that a UTA is a superposition of lines, a STA can be understood as a superposition of UTAs. The $n$-order $\mathcal{M}_n^{\ab}$ moments associated with a given $\alpha\rightarrow\beta$ mono-electronic jump are not normalized in this article. This explains the notion of 0th-order moment, which precisely represents this norm. Let us define
\begin{equation}
\mathcal{M}_n^{\ab}=\sum_{c\in\Xi}\left(E_c^{\ab}\right)^nG_ce^{-\beta(E_c^{\ab}-\mu Q)}=\langle E_c^n\rangle~\mathcal{U}_Q^{(\Xi)},
\end{equation}
where the energy of a configuration $c$ reads
\begin{equation}
E_c^{\ab}=D_0^{\alpha\beta}+\sum_r\left(p_r-\delta_{r,\alpha}\right)D_r^{\alpha\beta},
\end{equation}
and $\mathcal{U}_Q^{(\Xi)}$ is the partition function of superconfiguration $\Xi$:
\begin{equation}
\mathcal{U}_Q^{(\Xi)}=\sum_{c\in\Xi}G_ce^{-\beta(E_c-\mu Q)}.
\end{equation}
We can expand the zero- and first-order moments as
\begin{equation}
\mathcal{M}_0^{\ab}=\ga \gb X_\alpha \mathcal{U}_{Q-1}^\alpha\left(g^{\alpha\beta}\right)
\end{equation}
and
\begin{equation}
\mathcal{M}_1^{\ab}=\ga \gb X_\alpha \left[ \DO \mathcal{U}_{Q-1}^\alpha\left(g^{\alpha\beta}\right)
+\sum_r \Dr X_r g_r^{\alpha\beta} \mathcal{U}_{Q-2}^{\alpha r}\left(g^{\alpha\beta r}\right) \right].
\end{equation}
The functions $\DO=\langle \beta\rangle-\langle\alpha\rangle$ and $\Dr=2[\langle \beta,r\rangle-\langle \alpha,r\rangle]+\left(\delta_{r,\alpha}-\delta_{r,\beta}\right)\eta_r^{\alpha\beta}$ depend on the mono- $\langle r\rangle$ and di-electronic $\langle r,s\rangle$, as well as functions $\eta_r^{\alpha\beta}$ for which expressions can be found in Refs. \cite{Barshalom1989,Blenski1997,Blenski2000}. The notation $\mathcal{U}_{Q-1}^\alpha\left(g^{\alpha\beta}\right)$ means that

\begin{itemize}

\item i) The partition function is evaluated with $Q_\sigma-1$ electrons in the supershell $\sigma$ containing subshell $\alpha$, and $Q_\sigma$ in the others (with $\sum_\sigma Q_\sigma=Q$).

\item ii) The degeneracies of both subshells $\alpha$ and $\beta$ are reduced by one.

\end{itemize}

We set $g_r^{\alpha\beta}=g_r-\delta_{r,\alpha}-\delta_{r,\beta}$. The average energy of a STA is built from these two moments by the formula
\begin{eqnarray}
E_S^{\ab}&=&\frac{\mathcal{M}_1^{\ab}}{\mathcal{M}_0^{\ab}}=\DO + \sum_r \Dr X_r g_r^{\alpha\beta} \frac{\mathcal{U}_{Q-2}^{\alpha r}\left(g^{\alpha\beta r}\right)}
{\mathcal{U}_{Q-1}^{\alpha}\left(g^{\alpha\beta}\right)},
\end{eqnarray}
while the second-order can be split into five parts:

\begin{equation}
\mathcal{M}_2^{\ab}=\mathcal{M}_{21}^{\ab}+\mathcal{M}_{22}^{\ab}+\mathcal{M}_{23}^{\ab}+\mathcal{M}_{24}^{\ab}+\mathcal{M}_{25}^{\ab}.
\end{equation}
For the initial and final superconfigurations, a given single-electron $\ab$ transition can occur between several pairs of electronic configurations. Each of these $\ab$ lines has a slightly different energy, and its own UTA width. The $M_{21}^{\ab}$ moment is the contribution to the second-order moment of the UTA widths of each of these lines \cite{Barshalom1995}:

\begin{equation}
\mathcal{M}_{21}^{\ab}=\ga \gb X_\alpha \sum_r \mathcal{V}_G(\ell_r \ell_\alpha - \ell_r \ell_\beta) X_r g_r^{\alpha\beta}
\mathcal{U}_{Q-2}^{\alpha r}\left(g^{\alpha\beta rr}\right),
\end{equation}
where $\mathcal{V}_G(\ell_r \ell_\alpha - \ell_r \ell_\beta)$ is the UTA variance of the $\ab$ transition with a single spectator electron in the $r$ subshell. Analytical formulas for these variances are given in Ref. \cite{Bauche1979}. The second component represents the contribution of the spin-orbit interaction
\begin{equation}
\mathcal{M}_{22}^{\ab}=\Delta_{so,\ab}^2 \mathcal{M}_0^{\ab},
\end{equation}
with
\begin{equation}
\Delta_{so,\ab}^2=\frac{(\zeta_\alpha-\zeta_\beta)}{4}[\ell_\alpha(\ell_\alpha+1)\zeta_\alpha-\ell_\beta(\ell_\beta+1)\zeta_\beta]+\frac{\zeta_\alpha\zeta_\beta}{2},
\end{equation}
where $\zeta_r$ represents the usual spin-orbit integral. The remaining terms correspond to the energy spread of the lines of transition $\ab$ between two configurations:
\begin{eqnarray}
\mathcal{M}_{23}^{\ab}=\ga\gb X_\alpha\left[\sum_{r,t}\Dr\Dt g_t^{\alpha\beta r} g_r^{\alpha\beta}X_rX_t\mathcal{U}_{Q-3}^{\alpha rt}\left(g^{\alpha\beta rt}\right)+\sum_t\left(\Dt\right)^2 g_t^{\alpha\beta}X_t\mathcal{U}_{Q-2}^{\alpha t}\left(g^{\alpha\beta t}\right)\right],
\end{eqnarray}
\begin{equation}
\mathcal{M}_{24}^{\ab}=2 \DO \left( \mathcal{M}_1^{\ab} - \DO \mathcal{M}_0^{\ab} \right)
\end{equation}
and
\begin{equation}
\mathcal{M}_{25}^{\ab}=(\DO)^2 \mathcal{M}_0^{\ab}.
\end{equation}
The STA variance is finally
\begin{eqnarray}
\left(\Delta_S^{\ab}\right)^2=\frac{\mathcal{M}_2^{\ab}}{\mathcal{M}_0^{\ab}}-\left(\frac{\mathcal{M}_1^{\ab}}{\mathcal{M}_0^{\ab}}\right)^2.
\end{eqnarray}
The calculation of partition functions, involving modified degeneracies, may seem problematic. When the initial or final subshell of the mono-electronic transition is an ``s'' or ``p$_{1/2}$'' subshell (degeneracy $g=2$), the preceding formulae can introduce zero or negative degeneracies, making the formula indeterminate \textit{a priori}. For example, the partition function $\mathcal{U}_Q\left(g^{\alpha\beta r s}\right)$ makes no sense if $r=s=\alpha$ (or $r=s=\beta$) and $\alpha$ (or $\beta$) is an ``s'' subshell, since this means that the degeneracy of this orbital is equal to -1... Usually, this problem is circumvented by expressing partition functions using the following relation:
\begin{eqnarray}
\mathcal{U}_Q\left(g^{\alpha\beta r s}\right)&=&\sum_{i=0}^Q (-X_s)^i \mathcal{U}_{Q-i}\left(g^{\alpha\beta r}\right)=\sum_{i=0}^Q \sum_{j=0}^{Q-i}(-X_s)^i(-X_r)^j \mathcal{U}_{Q-i-j}\left(g^{\alpha\beta}\right)\nonumber\\
&=&\sum_{k=0}^Q \sum_{i=0}^{k}(-X_s)^i(-X_r)^{k-i} \mathcal{U}_{Q-k}\left(g^{\alpha\beta}\right)=\sum_{k=0}^Q (-X_r)^k \mathcal{U}_{Q-k}\left(g^{\alpha\beta}\right) \sum_{i=0}^{k}\left(\frac{X_s}{X_r}\right)^i.
\end{eqnarray}

Proceeding that way, there is no more ``apparent'' issue, since the functions $\mathcal{U}_{Q-k}\left(g^{\alpha\beta}\right)$ are always defined. However, a doubt remains, concerning the general validity of the final formulas, since they may rely on an indeterminate form at a given step of the calculation. The purpose of the next sections is to remove indeterminate forms, when they arise, by analyzing the different steps of the calculation of moments, and to check the impact it may have on the formulas published by Bar-Shalom \textit{et al.} \cite{Barshalom1989} for the STA variances. By examining the formulas for moments, we can see that two contributions are likely to be problematic: $\mathcal{M}_{23}^{\ab}$ and $\mathcal{M}_{21}^{\ab}$.


\subsection{Contribution $\mathcal{M}_{23}^{\ab}$}\label{subsec23}

As concerns the $\mathcal{M}_{23}^{\ab}$ part
\begin{eqnarray}\label{m23}
\mathcal{M}_{23}^{\ab}=\ga \gb X_\alpha \left[ \sum_{r,t} \Dr \Dt g_t^{\alpha\beta r} g_r^{\alpha\beta} X_r X_t
\mathcal{U}_{Q-3}^{\alpha rt}\left(g^{\alpha\beta rt}\right)+\sum_t \left(\Dt\right)^2 g_t^{\alpha\beta}X_t \mathcal{U}_{Q-2}^{\alpha
t}\left(g^{\alpha\beta t}\right)\right],
\end{eqnarray}
an indetermination may appear in the double sum for $r=t=\alpha$ (or $r=t=\beta$) if $\alpha$ (or $\beta$) is a ``s'' orbital: we must then calculate functions $\mathcal{U}_{Q-3}^{\alpha\alpha\alpha}\left(g^{\alpha\beta\alpha\alpha}\right)$ (or $\mathcal{U}_{Q-3}^{\alpha\beta\beta}\left(g^{\alpha\beta\beta\beta}\right)$) which have no \textit{a priori} meaning. However, we're saved by the term $g_t^{\alpha\beta r}$ which is zero in these situations. The formula (\ref{m23}) is therefore correct, and can be used in all cases.
 
\subsection{Contribution $\mathcal{M}_{21}^{\ab}$}\label{subsec24} 

An indetermination can arise due to the calculation of the UTA width:
\begin{equation}\label{m21a}
\mathcal{M}_{21}^{\ab}=\ga \gb X_\alpha \sum_r \mathcal{V}_G\left(\ell_r\ell_{\alpha}-\ell_r\ell_{\beta}\right) g_r^{\alpha\beta} X_r
\mathcal{U}_{Q-2}^{\alpha r}\left(g^{\alpha\beta rr}\right),
\end{equation}
when $r=\alpha$ (or $r=\beta$) and $\alpha$ (or $\beta$) is an ``s'' or ``p$_{1/2}$'' subshell. On the contrary to the preceding case, there is no multiplicative factor likely to cancel these contributions, and one is faced with the evaluation of quantities like $\mathcal{U}_{Q-2}^{\alpha\alpha}\left(g^{\alpha\beta\alpha\alpha}\right)$ (or $\mathcal{U}_{Q-2}^{\alpha\beta}\left(g^{\alpha\beta\beta\beta}\right)$) without clear physical meaning. The origin of such indeterminations stems from an incorrect manipulation in the derivation of the moment. Let us consider the mono-electronic transition $\ab$ connecting two configurations $c$ and $c'$ defined by
\begin{equation}
\left\{
\begin{tabular}{llllllll}
$c:$ & $\ell_1^{N_1}$ & $\ell_2^{N_2}$ & $\ldots$ & $\ell_{\alpha}^{N_\alpha}$ & $\ldots$ & $\ell_{\beta}^{N_\beta}$ & $\ldots$ \nonumber\\
$c':$ & $\ell_1^{N_1}$ & $\ell_2^{N_2}$ & $\ldots$ & $\ell_{\alpha}^{N_{\alpha-1}}$ & $\ldots$ & $\ell_{\beta}^{N_{\beta+1}}$ & $\ldots$
\end{tabular}
\right.
\end{equation}
The statistical broadening of the transition array $c-c'$ can be expressed as a combination of the two-electron variances $\vg(\ell_r \ell_\alpha-\ell_r \ell_\beta)$,
\begin{eqnarray}\label{rouge}
\vgab &=& \frac{p_1(g_1-p_1)}{g_1-1}\vg(\ell_1\ell_\alpha-\ell_1\ell_\beta) \\\nonumber
&+& \frac{p_2(g_2-p_2)}{g_2-1}\vg(\ell_2\ell_\alpha-\ell_2\ell_\beta) \\\nonumber
&+& \ldots\\\nonumber
&+& \frac{(p_\alpha-1)(g_\alpha-p_\alpha)}{g_\alpha-2}\vg(\ell_\alpha^2-\ell_\alpha\ell_\beta) \\\nonumber
&+& \ldots\\\nonumber
&+& \frac{p_\beta(g_\beta-p_\beta-1)}{g_\beta-2}\vg(\ell_\alpha\ell_\beta-\ell_\beta^2), \\\nonumber
&+& \ldots
\end{eqnarray}
each term corresponding to the accounting for potential spectator electrons in each subshell. The notation $\vec{p}=\left\{p_1, p_2, \cdots, p_N\right\}$ represents the vector of subshell populations. The latter expression can be put in the more compact form:
\begin{equation}\label{vgab}
\vgab=\sum_{s=1}^N \frac{\left(p_s-\delta_{s,\alpha}\right)\left(g_s-p_s-\delta_{s,\beta}\right)}{g_s^{\alpha\beta}-1}\vgsab,
\end{equation}
with the simplified notation $\vgsab=\mathcal{V}_G\left(\ell_r\ell_{\alpha}-\ell_r\ell_{\beta}\right)$. The moment $\mathcal{M}_{21}^{\ab}$ can be obtained by summing that variance over all the initial configurations of the superconfiguration:
\begin{equation}\label{m21b}
\mathcal{M}_{21}^{\ab}=\sum_{|\vec{p}|=Q}\prod_{j=1}^N \bin{g_j}{p_j}X_j^{p_j}\vgab p_\alpha (g_\beta-p_\beta),
\end{equation}
with $|\vec{p}|=p_1+p_2+\cdots+p_N$. Introducing expression (\ref{vgab}) into Eq. (\ref{m21b}), we arrive at expression (\ref{m21a}). The details of such calculations are provided in Ref. \cite{Barshalom1989}. The validity of such a formula is clear as soon as $g_\alpha\ne2$ and $g_\beta\ne2$ (note that these conditions can not apply simultaneously since $\alpha\ne\beta$). But if one of those conditions is verified, a division by zero, or an indeterminate form (zero divided by zero) arise in the expression of the moment. The formula (\ref{m21a}) may therefore deserve scrutiny in such situations.

\subsection{Case where $\alpha$ or $\beta$ is an ``s'' or ``p$_{1/2}$'' subshell}\label{subsec25}

\subsubsection{Case where $\alpha$ is a $s$ or $p_{1/2}$ subshell}\label{subsubsec251}

Let us consider the case where the initial subshell $\alpha$ is a $s$ subshell, \emph{i.e.}, $\ell_\alpha=0$ and $g_\alpha=4 \ell_\alpha+2=2$ (the case of the relativistic orbital $p_{1/2}$ is identical). In this case, there are two situations to consider: 

\begin{itemize}
\item if $p_{\alpha}=1$, there is no spectator in the orbital, since the single electron participates in the optical transition; 

\item if $p_{\alpha}=2$, we have a full ``s'' or ``p$_{1/2}$'' subshell, and therefore makes no contribution to the variance. 

\end{itemize}

Note that the case $p_\alpha=0$ is not allowed. There is therefore no contribution to variance from the term $\mathcal{V}_G\left(\ell_r\ell_{\alpha}-\ell_r\ell_{\beta}\right)$ which is zero by the way. If we calculate the variance of the transition array s$^2\rightarrow$ sp using the UTA formulas \cite{Bauche1979}, we find an indeterminate form with divisions by zero (because of $\ell=0$). In this case, we use the fact that this variance is identical for the complementary transition array $p^6\rightarrow p^5s$, and show that the variance is zero. Thus, the UTA variance is written in the following form (see the third line in the right-hand side):
\begin{eqnarray}\label{bleu}
\vgab &=& \frac{p_1(g_1-p_1)}{g_1-1}\vg(\ell_1\ell_\alpha-\ell_1\ell_\beta) \\\nonumber
&+& \frac{p_2(g_2-p_2)}{g_2-1}\vg(\ell_2\ell_\alpha-\ell_2\ell_\beta)+\ldots\\\nonumber
&+& 0\times\vg(\ell_\alpha^2-\ell_\alpha\ell_\beta)+\ldots\\\nonumber
&+& \frac{p_\beta(g_\beta-p_\beta-1)}{g_\beta-2}\vg(\ell_\alpha\ell_\beta-\ell_\beta^2)+\ldots
\end{eqnarray}
The latter formula can be put in the more compact form:
\begin{equation}\label{vgab2}
\vgab=\sum_{s\ne\alpha}\frac{\left(p_s-\delta_{s,\alpha}\right)\left(g_s-p_s-\delta_{s,\beta}\right)}{g_s^{\alpha\beta}-1}\vgsab.
\end{equation}
It is easy to check that the corresponding moment is equal to
\begin{equation}\label{m21a2}
\mathcal{M}_{21}^{\ab}=\ga \gb X_\alpha \sum_{r\ne\alpha} \vgrab g_r^{\alpha\beta} X_r\mathcal{U}_{Q-2}^{\alpha r}\left(g^{\alpha\beta rr}\right).
\end{equation}
It is precisely the latter relation that must be used {\it in lieu} of Eq. (\ref{m21a}) when $\alpha$ is a ``s'' subshell. It is useful to calculate the difference between Eqs. (\ref{m21a2}) and (\ref{m21a}), in order to quantify the potential error due to the ``original'' formulas; one has
\begin{eqnarray}
\mathcal{M}_{21}^{\ab}&=&
\ga \gb X_\alpha \sum_r \vgrab g_r^{\alpha\beta} X_r\mathcal{U}_{Q-2}^{\alpha r}\left(g^{\alpha\beta rr}\right)-
\ga \gb X_\alpha^2 \vgaab \mathcal{U}_{Q-2}^{\alpha\alpha}\left(g^{\alpha\beta\alpha\alpha}\right),
\end{eqnarray}
and thus
\begin{eqnarray}
\delta \mathcal{M}_{21}^{\ab}=\mathcal{M}_{21}^{\ab}-\left(\mathcal{M}_{21}^{\ab}\right)_\text{orig}=-\ga \gb X_\alpha^2 \vgaab \mathcal{U}_{Q-2}^{\alpha\alpha}\left(g^{\alpha\beta\alpha\alpha}\right).
\end{eqnarray}
As concerns the impact on the STA variance, one has
\begin{eqnarray}
\left(\delta\Delta_S^{\ab}\right)^2=\frac{\delta \mathcal{M}_{21}^{\ab}}{\mathcal{M}_{0}^{\ab}}=-\vgaab \frac{X_\alpha \mathcal{U}_{Q-2}^{\alpha\alpha}\left(g^{\alpha\beta\alpha\alpha}\right)}{\mathcal{U}_{Q-1}^{\alpha}\left(g^{\alpha\beta}\right)}.
\end{eqnarray}
Using the following relation
\begin{eqnarray}
X_\alpha \mathcal{U}_{Q-2}^{\alpha\alpha}\left(g^{\alpha\beta\alpha\alpha}\right)=\mathcal{U}_{Q-1}^{\alpha}\left(g^{\alpha\beta\alpha}\right)-\mathcal{U}_{Q-1}^{\alpha}\left(g^{\alpha\beta\alpha\alpha}\right),
\end{eqnarray}
we get
\begin{eqnarray}
\left(\delta\Delta_S^{\ab}\right)^2=
\vgaab \left(
\frac{\mathcal{U}_{Q-1}^{\alpha}\left(g^{\alpha\beta\alpha\alpha}\right)}{\mathcal{U}_{Q-1}^{\alpha}\left(g^{\alpha\beta}\right)}-
\frac{\mathcal{U}_{Q-1}^{\alpha}\left(g^{\alpha\beta\alpha}\right)}{\mathcal{U}_{Q-1}^{\alpha}\left(g^{\alpha\beta}\right)}
\right).
\end{eqnarray}
Let us now introduce the usual notation $\mathcal{U}_{Q-1}^{\alpha}=\mathcal{U}_{Q'}$, with 
\begin{equation}
Q'=\left\{
\begin{tabular}{ll}
$Q_\sigma-1$ & $\alpha\in\sigma$\\
$Q_\sigma$ & $\alpha\notin\sigma$
\end{tabular}\right..
\end{equation}
One has
\begin{eqnarray}
\mathcal{U}_{Q'}\left(g^{\alpha\beta\alpha}\right)=\sum_{k=0}^{Q'}(-X_\alpha)^k\:\mathcal{U}_{Q'-k}\left(g^{\alpha\beta}\right)
\end{eqnarray}
and
\begin{eqnarray}
\mathcal{U}_{Q'}\left(g^{\alpha\beta\alpha\alpha}\right)&=&\sum_{i=0}^{Q'}(-X_\alpha)^i\:\mathcal{U}_{Q'-i}\left(g^{\alpha\beta\alpha}\right)=\sum_{i=0}^{Q'}\sum_{j=0}^{Q'-i}(-X_\alpha)^{i+j}\:\mathcal{U}_{Q'-i-j}\left(g^{\alpha\beta}\right)\nonumber\\
&=&\sum_{i=0}^{Q'}\sum_{k=i}^{Q'}\left(-X_\alpha\right)^{k}\:\mathcal{U}_{Q'-k}\left(g^{\alpha\beta}\right)=\sum_{k=0}^{Q'}\sum_{i=0}^{k}\left(-X_\alpha\right)^{k}\:\mathcal{U}_{Q'-k}\left(g^{\alpha\beta}\right)\nonumber\\
&=&\sum_{k=0}^{Q'}(k+1)(-X_\alpha)^{k}\:\mathcal{U}_{Q'-k}\left(g^{\alpha\beta}\right).
\end{eqnarray}
Finally, we obtain the correction - to be added to the original formulas - in case where the initial subshell $\alpha$ of the radiative transition is an ``s'' or ``p$_{1/2}$'' subshell:
\begin{eqnarray}
\left(\delta\Delta_S^{\ab}\right)^2=\vgaab \sum_{k=0}^{Q'}k\left(-X_\alpha\right)^k \:\frac{\mathcal{U}_{Q'-k}\left(g^{\alpha\beta}\right)}{\mathcal{U}_{Q'}\left(g^{\alpha\beta}\right)}.
\end{eqnarray}

\subsubsection{Case where $\beta$ is an ``s'' or ``p$_{1/2}$'' subshell}\label{subsubsec252}

Let us now see what happens if $\beta$, the final subshell of the mono-electronic transition, is a $s$ subshell, \emph{i.e.} $\ell_\beta=0$ and $g_\beta=4 \ell_\beta+2=2$. Similarly to the preceding case (see section \ref{subsubsec251}), the UTA broadening reads
\begin{equation}\label{vgab3}
\vgab=\sum_{s\ne\beta}\frac{\left(p_s-\delta_{s,\alpha}\right)\left(g_s-p_s-\delta_{s,\beta}\right)}{g_s^{\alpha\beta}-1}\vgsab,
\end{equation}
leading to the following expression of the moment:
\begin{equation}\label{m21c}
\mathcal{M}_{21}^{\ab}=\ga \gb X_\alpha \sum_{r\ne\beta} \vgrab g_r^{\alpha\beta} X_r\mathcal{U}_{Q-2}^{\alpha r}\left(g^{\alpha\beta rr}\right).
\end{equation}
The difference with the original formula for the second-order moment reads
\begin{eqnarray}
\delta \mathcal{M}_{21}^{\ab}=\mathcal{M}_{21}^{\ab}-\left(\mathcal{M}_{21}^{\ab}\right)_\text{orig}=-g_\alpha g_\beta X_\alpha X_\beta \vgbab \mathcal{U}_{Q-2}^{\alpha\beta}\left(g^{\alpha\beta\beta\beta}\right).
\end{eqnarray}
As concerns the STA variances, the difference is
\begin{eqnarray}
\left(\delta\Delta_S^{\ab}\right)^2=\frac{\delta M_{21}^{\ab}}{M_{0}^{\ab}}=
-\vgbab \frac{X_\beta \mathcal{U}_{Q-2}^{\alpha\beta}\left(g^{\alpha\beta\beta\beta}\right)}
{\mathcal{U}_{Q-1}^{\alpha}\left(g^{\alpha\beta}\right)}.
\end{eqnarray}
Finally, the correction - to be added to the original formulas - in case where the final subshell $\beta$ of the mono-electronic transition is an ``s'' or ``p$_{1/2}$'' subshell:
\begin{eqnarray}
\left(\delta\Delta_S^{\ab}\right)^2=
\vgbab \sum_{k=1}^{Q'}k(-X_\beta)^k \:\frac{\mathcal{U}_{Q'-k}\left(g^{\alpha\beta}\right)}{\mathcal{U}_{Q'}\left(g^{\alpha\beta}\right)}.
\end{eqnarray}

\subsubsection{Conclusion concerning the validity of STA formulas}\label{subsubsec253}

In the problematic cases, when $\alpha$ or $ \beta$ is an ``s'' (in the non-relativistic case, $\ell=0$) or $\text{p}_{1/2}$ (in the relativistic case, $j=1/2$) subshell, we showed that the corrective term is proportional to $\mathcal{V}_G^{s\rightarrow sp}$ or to $\mathcal{V}_G^{\text{s}\rightarrow \text{ps}}$, \emph{i.e.} the variance of transition array $\text{s}^2\rightarrow \text{sp}$. Such a variance is equal to zero, since it involves an an ``s'' or ``p$_{1/2}$'' subshell which is full. In order to convince oneself, it is convenient to calculate the variance of the ``complementary'' array $\text{p}^6\rightarrow \text{p}^5\text{s}$. One has obviously $H_1=H_2=H_3=H_4=H_6=0$ (see Ref. \cite{Bauche1979} for the definitions of the $H_i$ quantities), since all the $3j$ symbols cancel. It is exactly the same for the $H_5$ term for which
\begin{equation}
H_5=9\left[\frac{1}{9}-\neufj{0}{1}{1}{1}{1}{0}{1}{0}{1}\right]\threej{0}{1}{1}{0}{0}{0}^4[G^{(1)}(\text{ps})]^2,
\end{equation}
where $G^{(1)}$ represents the exchange Slater integral of order 1 (see Eq. (\ref{subsec22})). Since
\begin{equation}
\neufj{0}{1}{1}{1}{1}{0}{1}{0}{1}=\frac{1}{9},
\end{equation}
one gets $H_5=0$, which completes
\begin{equation}
\vg (\text{s}^2-\text{sp})=\mathcal{V}_G\left(\text{p}^6-\text{p}^5\text{s}\right)=0.
\end{equation}
As a consequence, the original formulas for the STA moments are correct. The occurrence of ``indefinite'' partition functions is avoided by a prefactor equal to zero. Therefore, the STA moments up to order 2 in their most compact form (\emph{i.e.} involving the reduced degeneracies) can be easily evaluated with the help of our robust recursion relation.

\section{Fast computation of partition functions with reduced degeneracies}\label{sec3}

\subsection{Robust recursion relations}\label{subsec31}

One possibility consists in eliminating all the partition functions with reduced degeneracies in order to have only partition functions with full degeneracies. This enables one to save computation time, but remains problematic because of alternating signs. Using Eq. (\ref{troiquoi}) for an integer $0\leq n\leq Q$ gives
\begin{equation}\label{septcinq}
\mathcal{U}_{Q-n}\left(g^{ab}\right)=\sum_{p=0}^{Q-n}\mathcal{U}_{Q-n-p}(g)(-1)^{p}\sum_{k=0}^{p}X_a^{p-k}X_b^k
\end{equation}
or
\begin{equation}\label{septsix}
\mathcal{U}_{Q-n}\left(g^{ab}\right)=\sum_{k=n}^{Q}\mathcal{U}_{Q-k}(g)(-1)^{k-n}\sum_{p=0}^{k-n}X_a^{k-p-n}X_b^p.
\end{equation}
Equation (\ref{septcinq}) can be simplified as (see Eq. (\ref{troicinq})):
\begin{equation}
\mathcal{U}_{Q-n}\left(g^{ab}\right)=\sum_{p=0}^{Q-n}(-1)^{p}\left[\frac{X_b^{p+1}-X_a^{p+1}}{X_b-X_a}\right]\mathcal{U}_{Q-n-p}(g)
\end{equation}
and Eq. (\ref{septsix}) as
\begin{equation}
\mathcal{U}_{Q-n}\left(g^{ab}\right)=\sum_{k=n}^{Q}(-1)^{k-n}\left[\frac{X_b^{k-n+1}-X_a^{k-n+1}}{X_b-X_a}\right]\mathcal{U}_{Q-k}(g).
\end{equation}
In the particular case where $a=b$, we have therefore
\begin{equation}
\mathcal{U}_{Q-n}\left(g^{aa}\right)=\sum_{p=0}^{Q-n}\mathcal{U}_{Q-n-p}(g)(-1)^{p}X_a^p(p+1)
\end{equation}
or
\begin{equation}
\mathcal{U}_{Q-n}\left(g^{aa}\right)=\sum_{k=n}^{Q}\mathcal{U}_{Q-k}(g)(-1)^{k-n}X_a^{k-n}(k-n+1).
\end{equation}

In the first STA codes \cite{Barshalom1989,Blenski1997,Blenski2000}, the partition functions where obtained from the relation
\begin{equation}\label{barrel}
\mathcal{U}_Q=\frac{1}{Q}\sum_{k=1}^Q\chi_k\mathcal{U}_{Q-k},
\end{equation}
with
\begin{equation}
\chi_k=-\sum_{i=1}^Ng_i(-X_i)^k,
\end{equation}
$N$ being the number of subshells of the considered supershell. Introducing a dependence with respect to $\beta=1/(k_BT)$, Eq. (\ref{barrel}) can be expressed in the following form (with a scaled argument $k\beta$ for $\mathcal{U}_1$):
\begin{equation}\label{recast}
\mathcal{U}_Q(\beta)=\frac{1}{Q}\sum_{k=1}^Q\mathcal{U}_1(k\beta)\mathcal{U}_{Q-k}(\beta).
\end{equation}
The recurrence relation (\ref{barrel}) (or equivalently (\ref{recast})) is initialized with the condition $\mathcal{U}_0=1$. From the beginning, we have used the notation $\mathcal{U}_Q$, illustrating the fact that the partition function depends solely on the number of electrons. In fact, the partition function depends also on the number of subshells $N$ in the supershell. The partition function can thus be noted $\mathcal{U}_{Q,N}$. The relation (\ref{barrel}) is a sum of quantities with alternating signs, which are sources of numerical instabilities. The Wilson and Chen approach consists in working with ratios of consecutive partition functions \cite{Wilson1999}:
\begin{equation}
\mathcal{R}_Q=\frac{\mathcal{U}_Q}{\mathcal{U}_{Q-1}}
\end{equation}
yielding the nested form
\begin{equation}\label{rwil}
\mathcal{R}_Q=\frac{\chi_1}{Q}\left\{1+\frac{c_2}{\mathcal{R}_{Q-1}}\left\{1+\frac{c_3}{\mathcal{R}_{Q-2}}\left\{\vphantom{\frac{c_2}{\mathcal{R}_{Q-1}}}1+\cdots\left\{\vphantom{\frac{c_2}{\mathcal{R}_{Q-1}}}\right\}\right\}\right\}\right\},
\end{equation}
where
\begin{equation}
c_i=\frac{\chi_i}{\chi_{i-1}}.
\end{equation}
Equation (\ref{rwil}) can also be recast in a homothetic form
\begin{equation}
\frac{Q\mathcal{R}_Q(\beta)}{\mathcal{R}_1(\beta)}=1+\sum_{i=2}^Q(-1)^{i-1}\prod_{k=1}^{i-1}\frac{\mathcal{R}_1\left[(k+1)\beta\right]}{\mathcal{R}_1\left[k\beta\right]}\frac{1}{\mathcal{R}_{Q-k}(\beta)}.
\end{equation}
However, this clever idea only brought slight improvement, and the numerical instabilities, although less frequent, remain. This is the main reason why we proposed a doubly-recursive relation \cite{Gilleron2004}, both over the number of electrons $Q$ and the number of subshells $N$. The relation reads
\begin{equation}\label{gilpain}
\mathcal{U}_{Q,N}=\sum_{p_N=0}^{\min(Q,g_N)}\bin{g_N}{p_N}X_N^{p_N}\mathcal{U}_{Q-p_N,N-1},
\end{equation}
initialized by the condition $\mathcal{U}_{0,0}=1$. As explained in the introduction, the averages over the populations, required in the STA theory, involve partition function with reduced degeneracies. The form (\ref{pop3}) appears to be the most efficient one since it involves two partition functions $\mathcal{U}_Q\left(g^a\right)$ and $\mathcal{U}_{Q-1}\left(g^a\right)$ which are ``simultaneously'' (in the sense that they rely on the same set of degeneracies) obtained from the joint recursion relation (\ref{gilpain}), making the replacement $g\rightarrow g^a$.

The doubly-recursive relation from Ref. \cite{Gilleron2004} (see Eq. (\ref{gilpain})) is stable even for larger supershells that are almost completely filled, but is needlessly computationally expensive. In these cases, the generalization by Wilson \emph{et al.} starting from completely filled shells is more effective and as stable \cite{Wilson2007,Pain2020}:

\begin{equation}
\mathcal{U}_{G_N-h,N}=\sum_{m=\max\left(0,h-G_{N-1}\right)}^{\min(Q,g_N)}\bin{g_N}{m}X_N^m~\mathcal{U}_{G_{N-1}-h+m,N-1},
\end{equation}
where $h$ is the number of holes in the supershell. $G_N=G$ is the degeneracy of the supershell and $G_{N-1}$ represents the total degeneracy of the supershell minus the degeneracy of the subshell $N$. Note that the ``removed'' subshell can be any of the list, not necessarily the last one.

Figure \ref{brux2} displays the ratios of consecutive partition functions in two cases: supershell (3s3p3d) in a germanium plasma at $T=20$ eV and $\rho$=0.01 g/cm$^3$ (left), and supershell (3s3p3d4s4p4d4f) in an iron plasma at $T=30$ eV and $\rho$=0.01 g/cm$^3$ (right). In the former case, the recursion relations of Bar-Shalom \textit{et al.} \cite{Barshalom1989}, and the approach of Wilson and Chen using ratios of consecutive partition functions \cite{Wilson1999} can be applied, but only exploiting the particle-hole symmetry (they can not cover all the cases in electron counting). In the iron-plasma case, however, both previous methods fail, even within the electron-hole counting. Only the relation (\ref{gilpain}) gives the exact values.

Figures \ref{Fig2} and \ref{Fig3} represent respectively the average populations and population standard deviations (divided by the degeneracy) of all the subshells contained in the supershell (3s3p3d) in a copper plasma at $T=$100 eV and $\rho=$1 g/cm$^3$, as a function of the possible number of electrons in the supershell. We can see that the variation of the average population is different, according to the considered subshell. The curves are concave for 3s, 3p and 3d subshells, and convex for subshells of the N shell. However, inside a shell, the shapes are similar, although the values are different. For the standard deviations also, the M and N shells reveal a different trend, the maximum is for a number of electron smaller than the half-degeneracy in the case of 3s, 3p and 3d, and higher in the case of 4s, 4p, 4d and 4f. Figure \ref{Fig4} displays the correlation

\begin{equation}
    \langle p_ip_j\rangle-\langle p_i\rangle\langle p_j\rangle,
\end{equation}
where
\begin{equation}
    \langle p_ip_j\rangle=\frac{1}{\mathcal{U}_Q(g)}\sum_{\substack{\{p_s\}\\\sum_{s=1}^Np_s=Q}}p_ip_j\prod_{s=1}^N\bin{g_s}{p_s}X_s^{p_s},
\end{equation}
for pairs of subshells ($i$=3d, $j$=4s) and ($i$=3d, $j$=4d). The above mentioned normalized standard deviation is therefore
\begin{equation}
    \frac{1}{g_i}\sqrt{\langle p_i^2\rangle-\langle p_i\rangle^2}.
\end{equation}

\subsection{Fast approximation}\label{subsec32}

With the notation $\bar{g}^a$ meaning that the subshell $a$ is excluded from the vector of degeneracies of the supershell, Oreg derived the following expression \cite{Oreg1997}:
\begin{equation}\label{oreg}
\mathcal{U}_Q(\bar{g}^a)=\sum_{n_1=0}^Q\sum_{n_2=0}^{n_1}\cdots\sum_{n_{g_a}=0}^{n_{g_a-1}}(-X_a)^{Q-n_{g_a}}\mathcal{U}_{n_{g_a}}(g)=\sum_{k=0}^QN_k^{(g_a)}(-X_a)^{Q-k}\mathcal{U}_k(g),
\end{equation}
with $N_k^{(g_a)}$ the number of times $\mathcal{U}_k(g)$ appears in the sum over all indices $n_i$. It is clear \cite{Oreg1997} that the specific value $n_{g_a}=k$ appears exactly once for each set $n_1\geq n_2\geq n_3\geq\cdots\geq n_{g_a-1}\geq k$. Thus 
\begin{equation}\label{oregnot}
N_k^{(m)}=\sum_{n_1=k}^Q\sum_{n_2=k}^{n_1}\cdots\sum_{n_{m-1}=k}^{n_{m-2}}1=\sum_{n_{m-1}=0}^{Q-k}\sum_{n_{m-2}=0}^{n_{m-1}}\cdots\sum_{n_2=0}^{n_3}\sum_{n_1=0}^{n_2}1.
\end{equation}
Oreg tabulated these numbers using Bernoulli functions. However, as mentioned by Faussurier \cite{Faussurier1999}, those numbers (noted $S_m(Q)$ by Faussurier), are simply the number of combinations with repetitions of $k$ elements among $N$, \emph{i.e.},
\begin{equation}\label{88}
N_k^{(m)}=\bin{Q-k+m-1}{m-1}.
\end{equation}
Such a result can be proven by induction. We have, $N_k^{(m)}=S_m(Q-k)$, following Faussurier's notation
\begin{equation}
S_m(Q)=\sum_{n_{m-1}=0}^{Q}\sum_{n_{m-2}=0}^{n_{m-1}}\cdots\sum_{n_2=0}^{n_3}\sum_{n_1=0}^{n_2}1.
\end{equation}
Noting that, for $m>2$:
\begin{equation}
S_m(Q)=\sum_{n_{m-1}=0}^{Q}S_{m-1}\left(n_{m-1}\right),
\end{equation}
we have
\begin{equation}
S_m(Q)=\sum_{n_{m-1}=0}^{Q}\bin{n_{m-1}+m-2}{m-2}.
\end{equation}
The identity
\begin{equation}
\sum_{n=0}^Q\bin{n+p}{p}=\bin{Q+p+1}{p+1}
\end{equation}
then completes the proof by induction.

It is worth mentioning that (see Appendix \ref{appA}), when one tries to find an equivalent version of relation in an inverse way (\emph{i.e.} expressing the partition function for a given number of subshells in terms of the partition functions for higher numbers of subshells), we get a relation with alternating signs and a combination with repetition as well. 

\begin{table}
\begin{center}
\begin{tabular}{cccc}\hline
Orbital & $\varepsilon _{n\ell}$ (eV) & ${g_{n\ell}}$ & $\Delta_{n\ell}$\\\hline\hline
3s & -369.82378 &  2 & 3.5638435082145383 \\
3p & -326.10399 &  6  & 1.9475291499937191\\
3d & -260.22501 & 10 & 0.52527915095661570\\
4s & -117.83349 &  2 & -0.63275928872077869 \\
4p & -101.62248 &  6 & -0.68771775131188129\\
4d & -77.903611 & 10 & -0.75365851294335506\\
4f & -59.280040 & 14 & -0.79551737222936836\\\hline
\end{tabular}
\caption{\label{tab:table1}%
Supershell orbitals, energies (eV), degeneracies and quantities $\Delta_{n\ell}$ for a non-relativistic treatment of a copper plasma at a temperature of 100 eV and density of 1 g/cm$^3$ corresponding to a chemical potential of -402.85531 eV.}
\end{center}
\end{table}

\begin{table}
\begin{center}
\begin{tabular}{ccccccccc}\hline
$k\downarrow \ m\rightarrow$ & 1 & 2 & 3 & 4 & 5 & 6 & 7 & 8\\\hline\hline 
0 & 1 & 8 & 36 & 120 & 330 & 792 & 1716 & 3432\\
1 & 1 & 7 & 28 & 84 & 210 & 462 & 924 & 1716 \\
2 & 1 & 6 & 21 & 56 & 126 & 252 & 462 & 792 \\
3 & 1 & 5 & 15 & 35 & 70 & 126 & 210 & 330 \\
4 & 1 & 3 & 9 & 20 & 35 & 56 & 84 & 120 \\
5 & 1 & 2 & 5 & 9 & 15 & 21 & 28 & 36 \\
6 & 1 & 2 & 2 & 3 & 5 & 6 & 7 & 8 \\
7 & 1 & 1 & 1 & 1 & 1 & 1 & 1 & 1 \\\hline
\end{tabular}
\caption{\label{tab:table2}%
Numbers $N_k^{(m)}=\bin{Q-k+m-1}{m-1}$ (see Eq. (\ref{88}) for different values of $k$ and $m$ in the case $Q=7$).}
\end{center}
\end{table}

\begin{figure}
\begin{center}
\includegraphics[scale=0.5]{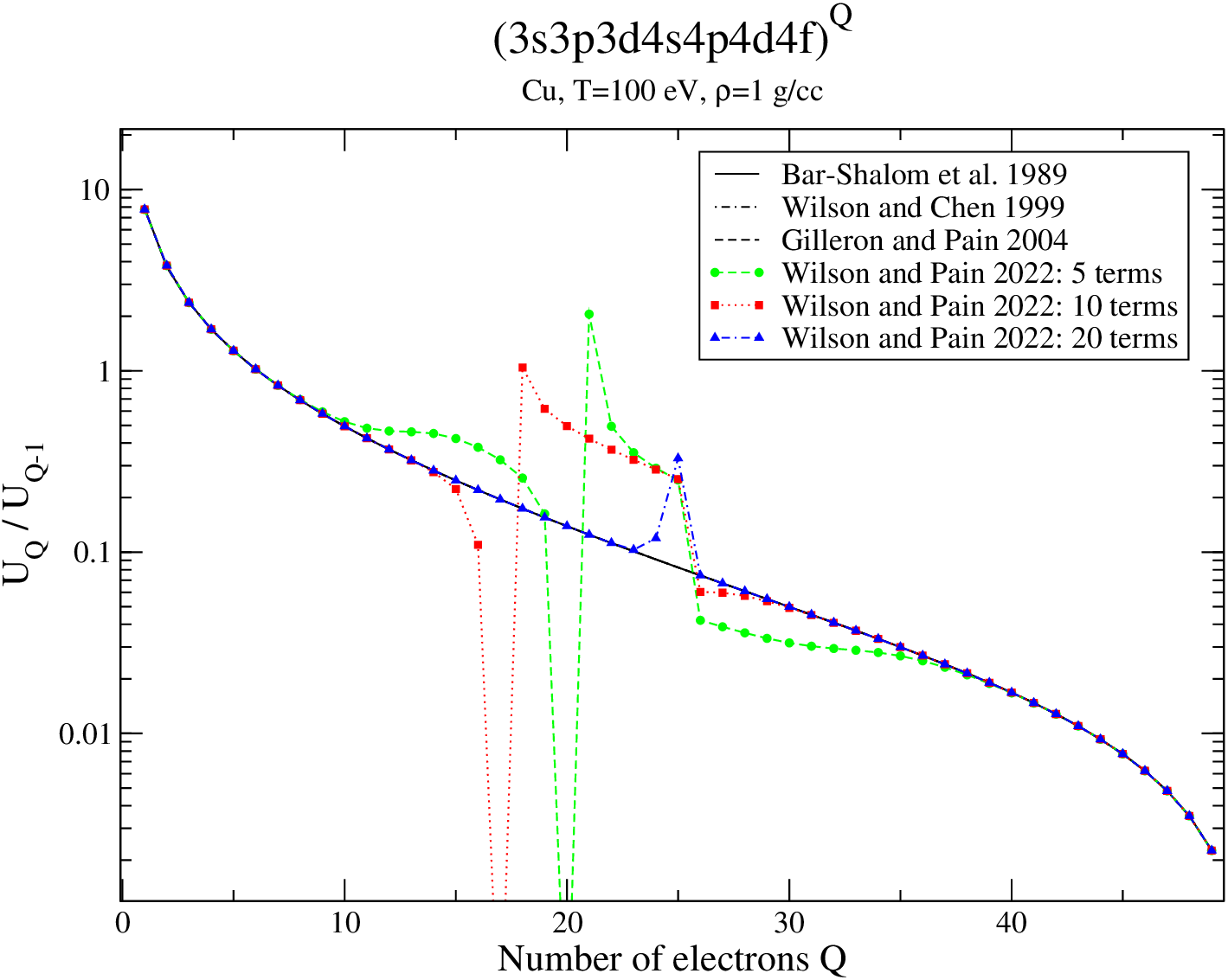}
\caption{Ratios of consecutive partition functions in supershell (3s3p3d4s4p4d4f) in a copper plasma at $T=$100 eV and $\rho=$1 g/cm$^3$ using the Wilson and Pain relation (\ref{wilexp}) \cite{Wilson2022} for different values of the number of terms in the expansion: 5, 10 and 20. Results of the recursion relations of Bar-Shalom \textit{et al.}, Wilson and Chen, and Gilleron and Pain (Eq. (\ref{gilpain})) are also displayed.}\label{Fig1}
\end{center}
\end{figure}

\begin{figure}[!ht]
 \begin{minipage}[c]{0.48\textwidth}
   \centering
   \includegraphics[width=\textwidth,angle=\anglefig,scale=\scalefig]{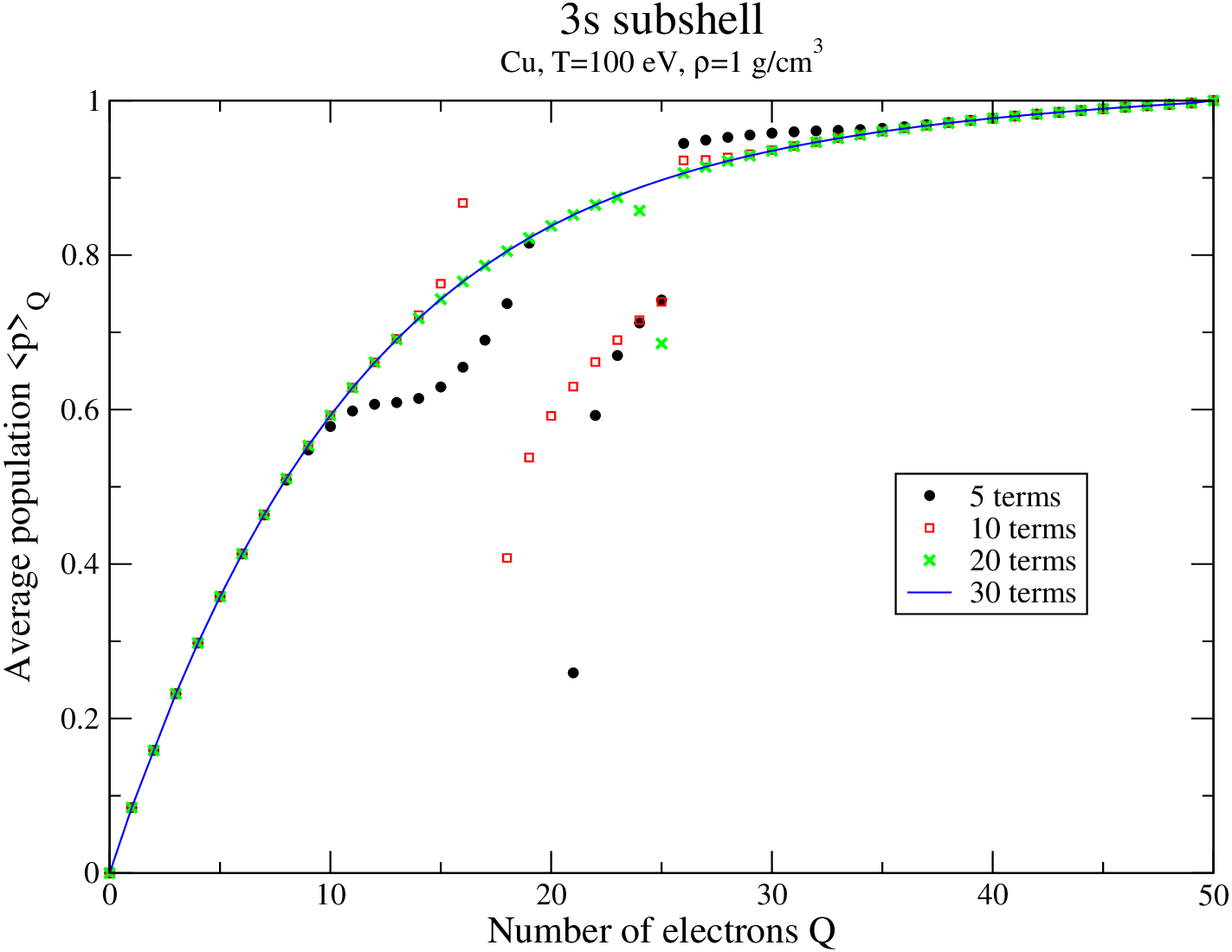}
   \caption{Population of subshell 3s in supershell (3s3p3d4s4p4d4f) for a copper plasma at $T=$100 eV and $\rho=$1 g/cm$^3$ using the relation (\ref{wilexp}) \cite{Wilson2022} for different values of the number of terms in the expansion: 10, 20 and 30.}\label{pop3s}
 \end{minipage}\hfill
 \begin{minipage}[c]{0.48\textwidth}
   \centering
   \includegraphics[width=\textwidth,angle=\anglefig,scale=\scalefig]{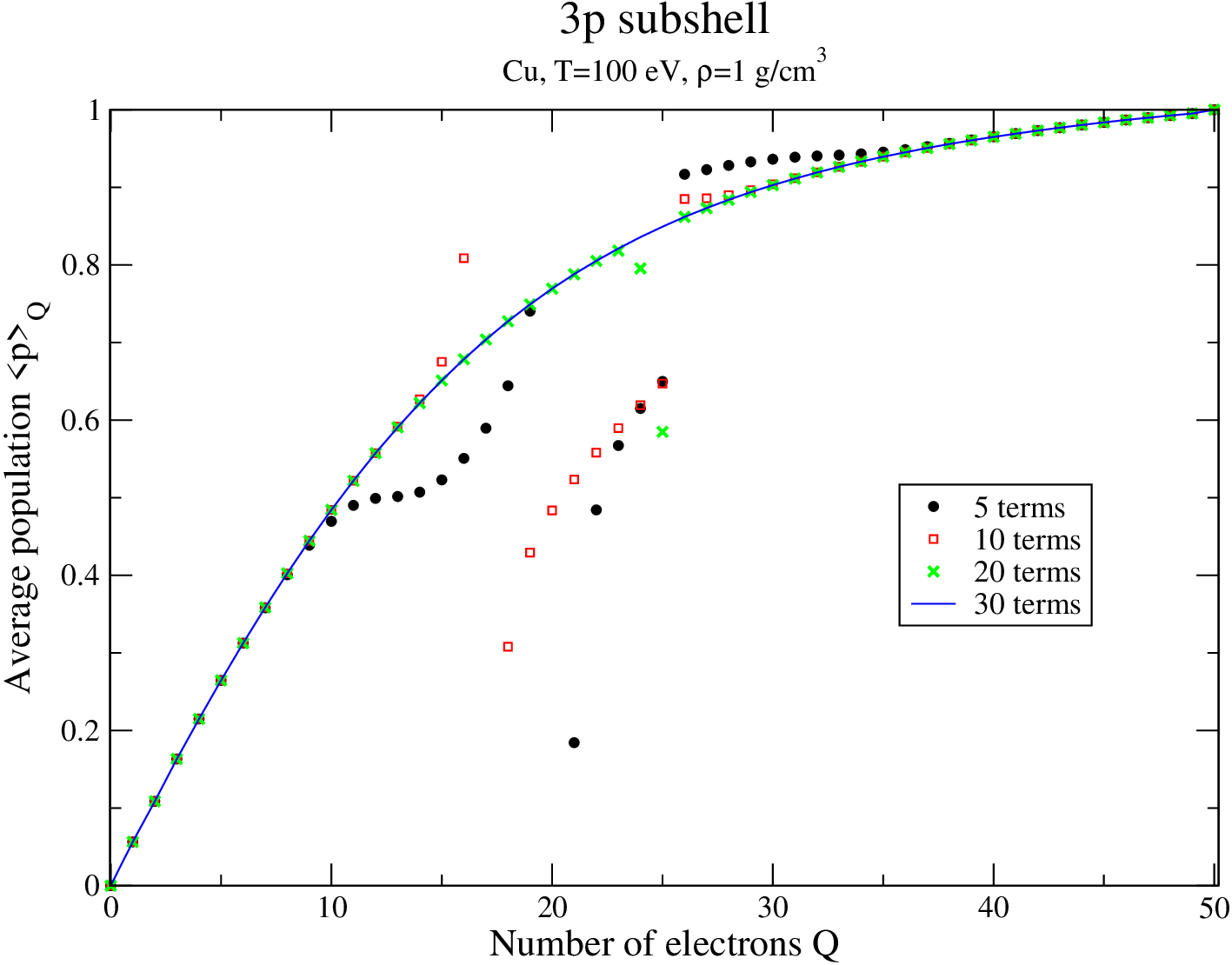}
   \caption{Population of subshell 3p in supershell (3s3p3d4s4p4d4f) for a copper plasma at $T=$100 eV and $\rho=$1 g/cm$^3$ using the relation (\ref{wilexp}) \cite{Wilson2022} for different values of the number of terms in the expansion: 10, 20 and 30.}\label{pop3p}
 \end{minipage}
\end{figure}

\begin{figure}[!ht]
 \begin{minipage}[c]{0.48\textwidth}
   \centering
   \includegraphics[width=\textwidth,angle=\anglefig,scale=\scalefig]{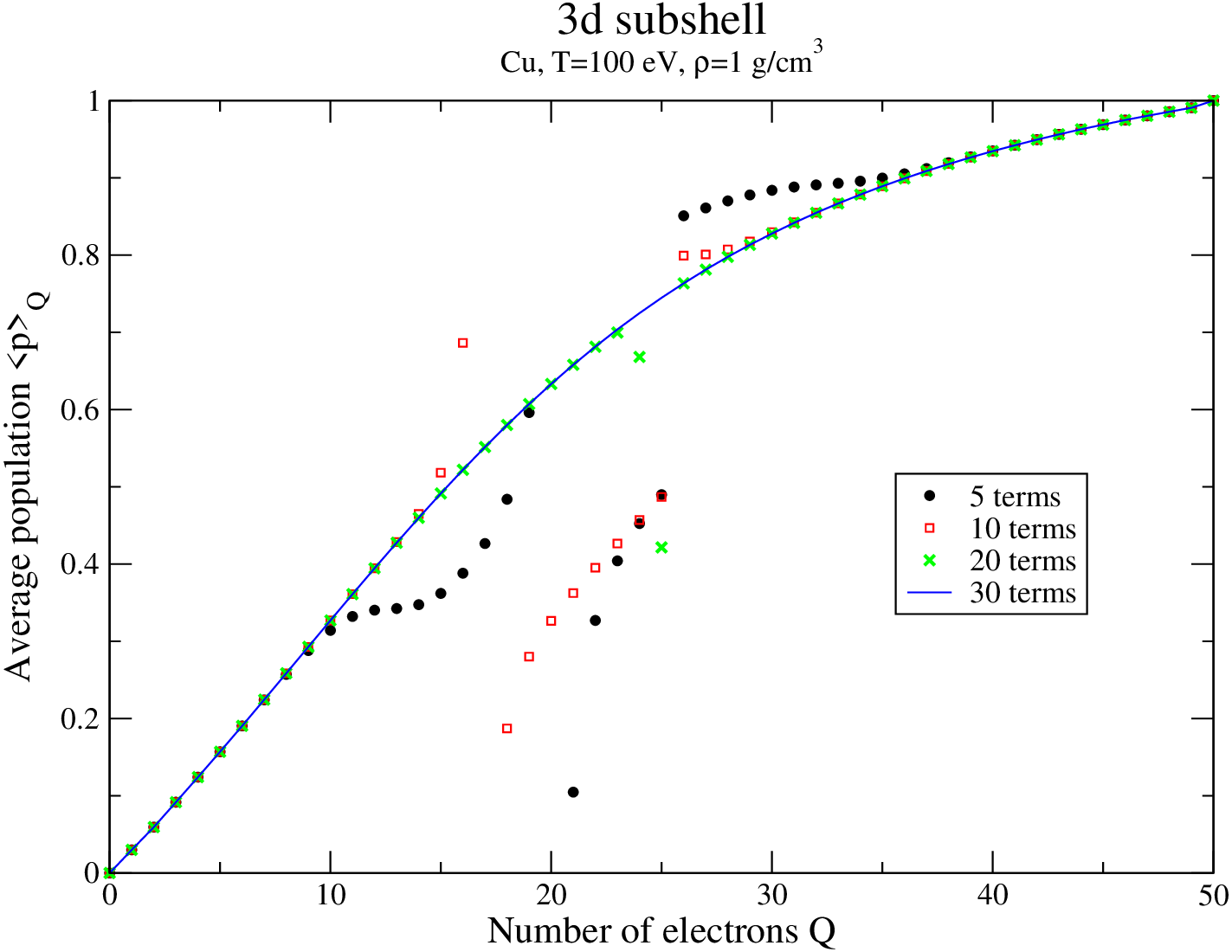}
   \caption{Population of subshell 3d in supershell (3s3p3d4s4p4d4f) for a copper plasma at $T=$100 eV and $\rho=$1 g/cm$^3$ using the relation (\ref{wilexp}) \cite{Wilson2022} for different values of the number of terms in the expansion: 10, 20 and 30.}\label{pop3d}
 \end{minipage}\hfill
 \begin{minipage}[c]{0.48\textwidth}
   \centering
   \includegraphics[width=\textwidth,angle=\anglefig,scale=\scalefig]{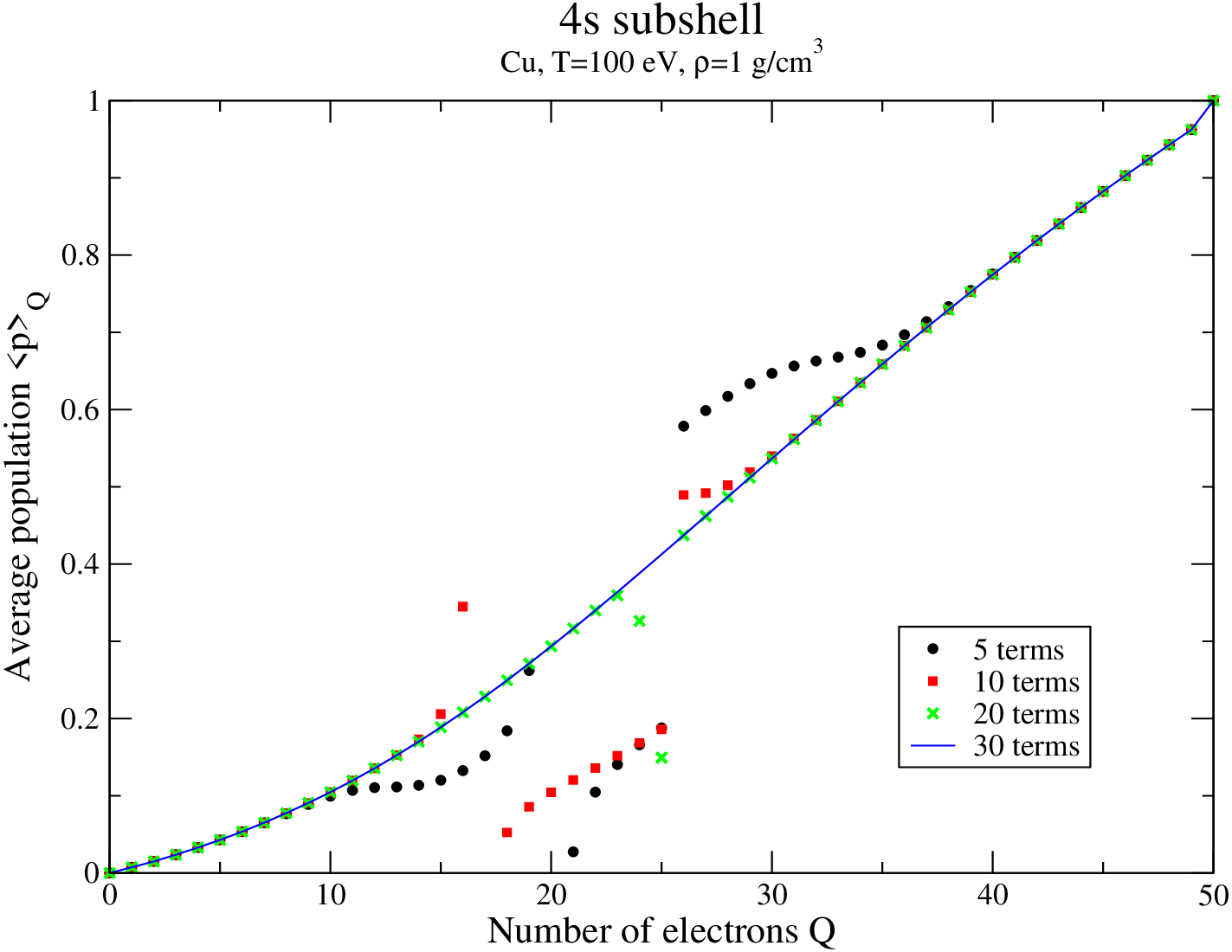}
   \caption{Population of subshell 4s in supershell (3s3p3d4s4p4d4f) for a copper plasma at $T=$100 eV and $\rho=$1 g/cm$^3$ using the relation (\ref{wilexp}) \cite{Wilson2022} for different values of the number of terms in the expansion: 10, 20 and 30.}\label{pop4s}
 \end{minipage}
\end{figure}

\begin{figure}
\begin{center}
\includegraphics[scale=0.35]{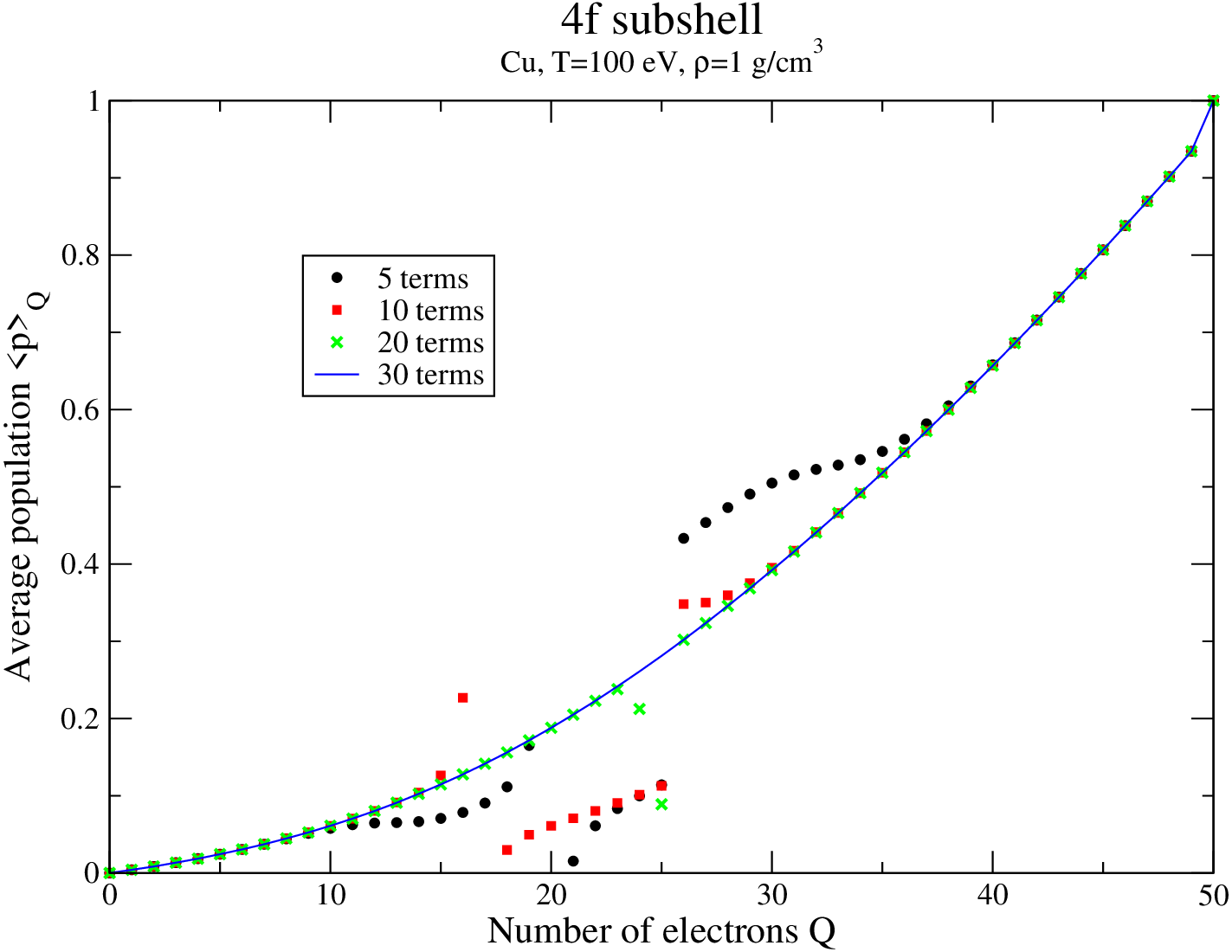}
\caption{Population of subshell 4f in supershell (3s3p3d4s4p4d4f) for a copper plasma at $T=$100 eV and $\rho=$1 g/cm$^3$ using the relation (\ref{wilexp}) \cite{Wilson2022} for different values of the number of terms in the expansion: 10, 20 and 30.}\label{pop4f}
\end{center}
\end{figure}

Figure \ref{Fig1} represents the ratios of consecutive partition functions in supershell (3s3p3d4s4p4d4f) for a copper plasma at $T=$100 eV and $\rho=$1 g/cm$^3$. Figures \ref{pop3s}, \ref{pop3p}, \ref{pop3d}, \ref{pop4s} and \ref{pop4f} represent the average populations of orbitals 3s, 3p, 3d, 4s and 4f respectively, as functions of the number of electrons $Q$. Inside a shell, the curves are very similar (see the behaviours of 3s, 3p and 3d on one hand, and of 4s, 4p, 4d and 4f on the other hand). This is the reason why we only show the 4s and 4f cases for the N shell. We can see that the convergence is not completely reached for 20 terms, the optimum value of terms being between 20 and 30 in all cases.

Recently, we published \cite{Wilson2022,Pain2023} a formula for supershell partition functions, which takes the form of a functional of the distribution of energies within the supershell and allows for fast and accurate computations, truncating the number of terms in the expansion. The latter involves coefficients $\Gamma_k$ for which we obtained a recursion relation and an explicit formula. One has
\begin{equation}\label{wilexp}
{\mathcal{U}_Q}(g) = X_0^Q\left\{ {\left( {\begin{array}{*{20}{c}}
G\\
Q
\end{array}} \right) + \left( {\begin{array}{*{20}{c}}
{G - 1}\\
{Q - 1}
\end{array}} \right){\Gamma_1} + \frac{1}{2}\left( {\begin{array}{*{20}{c}}
{G - 2}\\
{Q - 2}
\end{array}} \right){\Gamma_2} + \frac{1}{6}\left( {\begin{array}{*{20}{c}}
{G - 3}\\
{Q - 3}
\end{array}} \right){\Gamma_3} + \frac{1}{24}\left( {\begin{array}{*{20}{c}}
{G - 4}\\
{Q - 4}
\end{array}} \right){\Gamma_4} ...} \right\},
\end{equation}
where
\begin{equation}
{\Gamma_1} = \sum\limits_{i = 1}^m {{g_i}{\Delta _i}} 
\end{equation}
and
\begin{equation}
{\Gamma_2} = \left( {\sum\limits_{i = 1}^m {{g_i}{\Delta _i}} } \right)^2 - \left( {\sum\limits_{i = 1}^m {{g_i}\Delta _i^2} } \right),
\end{equation}
as well as
\begin{equation}
{\Gamma_3} = \left( {\sum\limits_{i = 1}^m {{g_i}{\Delta _i}} }\right)^3 -3\left( {\sum\limits_{j = 1}^m {{g_j}{\Delta _j}} } \right)\left( {\sum\limits_{i = 1}^m {{g_i}\Delta _i^2} } \right) +2 \left( {\sum\limits_{i = 1}^m {{g_i}\Delta _i^3} } \right)
\end{equation}
and also
\begin{eqnarray}\label{gamma4}
\Gamma_4&=&\left(\sum\limits_{i=1}^m g_i\Delta_i\right)^4-6\left(\sum\limits_{j=1}^mg_j\Delta_j\right)^2\left(\sum\limits_{i=1}^mg_i\Delta_i^2\right)+3\left(\sum\limits_{i=1}^mg_i\Delta _i^2\right)^2\nonumber\\
& &+8\left(\sum\limits_{i=1}^m g_i\Delta_i\right)\left(\sum\limits_{i=1}^mg_i\Delta_i^3\right)-6\left(\sum\limits_{i=1}^m g_i\Delta_i^4\right),
\end{eqnarray}
where
\begin{equation}
\Delta_i=\frac{X_i}{X_0}-1
\end{equation}
and
\begin{equation}
X_0=\frac{1}{G}\sum_{i=1}^Ng_iX_i.
\end{equation}
In the present case (copper plasma at $T$=100 eV and $\rho$=1 g/cm$^3$), we have $X_0=0.15747627875868209$ and the $\Delta_i$ are given in table \ref{tab:table1}.

In our previous work \cite{Wilson2022}, we found that the $\Gamma_k$ coefficients satisfy the recurrence relation
\begin{equation}\label{recur}
\left\{ \frac{\Gamma_k}{k!} \right\} = \frac{1}{k}\sum\limits_{p = 1}^k \left(-1\right)^{p+1}\Omega_p \,\left\{ \frac{\Gamma_{k - p}}{\left( k - p \right)!} \right\},
\end{equation}
with $\Gamma_0 = 1$ and
\begin{equation}
\Omega_p=\left[ {\sum\limits_{i = 1}^m {{g_i}\Delta _i^p} } \right].
\end{equation}
An explicit formula for the $\Gamma_k$ coefficients was obtained in Ref. \cite{Pain2023}, together with a mathematical proof based on elementary symmetric polynomials and the Newton-Girard identity. The numerical implementation is discussed in the same paper. The $\Gamma_k$ coefficients are actually equal to
\begin{equation}\label{newexpress}
\Gamma_k=k!(-1)^k\sum_{\substack{\vec{q}/\\q_1+2q_2+\cdots+kq_k=k}}\prod_{p=1}^k\frac{1}{q_p!}\left(-\frac{1}{p}\sum_{i=1}^mg_i\Delta_i^p\right)^{q_p}.
\end{equation}
Inserting expansion (\ref{wilexp}) into Eq. (\ref{oreg}) enables one to write
\begin{equation}
\mathcal{U}_Q\left(\bar{g}^{a}\right)=\sum_{k=0}^QX_0^k(-X_a)^{Q-k}\sum_{i=0}^{Q-k}\frac{\Gamma_i}{i!}\bin{G-i}{k-i}\bin{Q+k-1}{Q-1},
\end{equation}
which is equal to
\begin{equation}
\mathcal{U}_Q\left(\bar{g}^{a}\right)=\sum_{i=0}^{Q}\sum_{k=i}^QX_0^k(-X_a)^{Q-k}\frac{\Gamma_i}{i!}\bin{G-i}{k-i}\bin{Q+k-1}{Q-1}.
\end{equation}
Although it is probably of limited use for numerical applications, it is worth mentioning that the latter formula can be expressed in terms of hypergeometric functions:
\begin{eqnarray}
\mathcal{U}_Q\left(\bar{g}^{a}\right)&=&(-X_a)^Q\sum_{i=0}^{Q}\frac{\Gamma_i}{i!}\left\{\bin{Q+i-1}{Q-1}~_2F_1\left[
\begin{array}{c}
-G+i,Q+i\\
i+1\\
\end{array};\frac{X_0}{X_a}\right]\right.\nonumber\\
& &\left.-\bin{G-i}{Q+1-i}\bin{2Q}{Q-1}~_3F_2\left[
\begin{array}{c}
1,Q-G+1,2Q+1\\
Q+2,Q+2-i\\
\end{array};\frac{X_0}{X_a}\right]\right\},
\end{eqnarray}
where $~_2F_1$ and $~_3F_2$ represent the Gauss and Clausen hypergeometric functions respectively.

\section{Positive-definite expression of any $n^{th}$ order moment}\label{sec4}

\subsection{Derivation of the $n^{th}$ order moment neglecting UTA widths}\label{subsec41}

Neglecting the contribution of fine structure, \textit{i.e.} the UTA width, the unnormalized and uncentered $n^{th}-$order moment reads
\begin{equation}
\mathscr{M}_{n}^{\alpha\beta}=\sum_{|\vec{p}|=Q}p_{\alpha}\left(g_{\beta}-p_{\beta}\right)\left[D_0+\sum_{s=1}^N\left(p_s-\delta_{s, \alpha}\right)D_s\right]^n\prod_{k=1}^N\bin{g_k}{p_k}X_k^{p_k}.
\end{equation}
Since one has
\begin{equation}\label{identit}
p_{\alpha}\left(g_{\beta}-p_{\beta}\right)=g_{\alpha}g_{\beta}\frac{\bin{g_{\beta}-1}{p_{\beta}}}{\bin{g_{\beta}}{p_{\beta}}}-g_{\alpha}g_{\beta}\frac{\bin{g_{\alpha}-1}{p_{\alpha}}\bin{g_{\beta}-1}{p_{\beta}}}{\bin{g_{\alpha}}{p_{\alpha}}\bin{g_{\beta}}{p_{\beta}}},
\end{equation}
$\mathscr{M}_{n}^{\alpha\beta}$ can be put in the form
\begin{equation}
\mathscr{M}_n^{\alpha\beta}=\mathscr{M}_{n,1}^{\alpha\beta}-\mathscr{M}_{n,2}^{\alpha\beta}
\end{equation}
with
\begin{equation}
\mathscr{M}_{n,1}^{\alpha\beta}=g_{\alpha}g_{\beta}\sum_{|\vec{p}|=Q}\left[D_0+\sum_{s=1}^N\left(p_s-\delta_{s, \alpha}\right)D_s\right]^n\prod_{k=1}^N\bin{g_k-\delta_{k, \beta}}{p_k}X_k^{p_k} 
\end{equation}
and
\begin{equation}
\mathscr{M}_{n,2}^{\alpha\beta}=g_{\alpha}g_{\beta}\sum_{|\vec{p}|=Q}\left[D_0+\sum_{s=1}^N\left(p_s-\delta_{s, \alpha}\right)D_s\right]^n\prod_{k=1}^N\bin{g_k-\delta_{k, \alpha}-\delta_{k, \beta}}{p_k}X_k^{p_k}. 
\end{equation}
We have also
\begin{equation}
\left[D_0+\sum_{s=1}^N\left(p_s-\delta_{s, \alpha}\right)D_s\right]^n=\sum_{i=0}^n\bin{n}{i}\left(D_0-D_{\alpha}\right)^{n-i}\left(\sum_{s=1}^Np_sD_s\right)^i.
\end{equation}
The relation
\begin{equation}
\left(\sum_{p=0}^qa_p\right)^n=\sum_{k_0+k_1+\cdots+k_q=n}\bin{n}{k_0,k_1,\cdots,k_q}a_0^{k_0}a_1^{k_1}a_2^{k_2}\cdots a_q^{k_q}
\end{equation}
where
\begin{equation}
\bin{n}{k_0,k_1,\cdots,k_q}=\frac{n!}{k_0!k_1!\cdots k_q!}
\end{equation}
is the usual multinomial coefficient, yields
\begin{equation}
\left(\sum_{s=1}^Np_sD_s\right)^k=\sum_{i_1+i_2+\cdots+i_N=k}\bin{k}{i_1,i_2,\cdots,i_N}a_1^{i_1}a_2^{i_2}\cdots a_N^{i_N}
\end{equation}
with $a_s=p_sD_s$. One has therefore
\begin{equation}
\left(\sum_{s=1}^Np_sD_s\right)^k=\sum_{i_1+i_2+\cdots+i_N=k}\bin{k}{i_1,i_2,\cdots,i_N}D_1^{i_1}D_2^{i_2}\cdots D_N^{i_N}~p_1^{i_1}p_2^{i_2}\cdots p_N^{i_N}.
\end{equation}
We now take advantage of the relation
\begin{equation}\label{stir}
p^n\bin{g}{p}=\sum_{k=1}^nS_n^{(k)}(g)_k\bin{g-k}{p-k},
\end{equation}
with $S_n^{(k)}$ the Stirling numbers of the second kind and $(g)_k=g(g-1)\cdots(g-k+1)$. The latter Stirling numbers satisfy (among others) the relations
\begin{equation}
S_{n+1}^{(k)}=kS_n^{(k)}+S_n^{(k-1)}
\end{equation}
for $0 < k < n$ as well as
\begin{equation}
S_n^{(k)}={\frac {1}{k!}}\sum_{i=0}^{k}(-1)^{k-i}{\binom {k}{i}}i^{n}=\sum_{i=0}^{k}{\frac{(-1)^{k-i}i^{n}}{(k-i)!i!}},
\end{equation}
and
\begin{equation}
\sum_{n=0}^{\infty}S_n^{(k)}x^n=\prod_{p=1}^k\frac{x}{1-px}.
\end{equation}
One has in particular $S_n^{(n)}=1$, and $S_n^{(0)}=S_0^{(n)}=1$. Finally, using Eq. (\ref{stir}), we get
\begin{eqnarray}
p_1^{i_1}\cdots p_N^{i_N}\prod_{k=1}^N\bin{g_k-\delta_{k, \alpha}}{p_k}&=&\prod_{i=1}^Np_k^{i_k}\bin{g_k-\delta_{k, \alpha}}{p_k}\nonumber\\
&=&\left\{\sum_{r_1=1}^{i_1}S_{i_1}^{(r_1)}(g_1)_{r_1}\bin{g_1-\delta_{1, \alpha}-r_1}{p_1-r_1}\right\}\nonumber\\
& &\times\left\{\sum_{r_2=1}^{i_2}S_{i_2}^{(r_2)}(g_2)_{r_2}\bin{g_2-\delta_{2, \alpha}-r_2}{p_2-r_2}\right\}\cdots\nonumber\\
& &\times\left\{\sum_{r_N=1}^{i_N}S_{i_N}^{(r_N)}(g_N)_{r_N}\bin{g_N-\delta_{N, \alpha}-r_N}{p_N-r_N}\right\}\nonumber\\
& &=\sum_{r_1=1}^{i_1}\sum_{r_2=1}^{i_2}\cdots\sum_{r_N=1}^{i_N}\prod_{k=1}^NS_{i_k}^{(r_k)}(g_k)_{r_k}\bin{g_k-\delta_{k, \alpha}-r_k}{p_k-r_k}.
\end{eqnarray}
The final expression of the contribution $\mathscr{M}_{n,1}^{\alpha\beta}$ to the moment is therefore
\begin{eqnarray}\label{mn1}
\mathscr{M}_{n,1}^{\alpha\beta}&=&g_{\alpha}g_{\beta}\sum_{|\vec{p}|=Q}\sum_{i=0}^n\bin{n}{i}\left(D_0-D_{\alpha}\right)^{n-i}\sum_{i_1+i_2+\cdots+i_N=i}\bin{i}{i_1,i_2,\cdots,i_N}\nonumber\\
& &\times D_1^{i_1}D_2^{i_2}\cdots D_N^{i_N}\sum_{r_1=1}^{i_1}\sum_{r_2=1}^{i_2}\cdots\sum_{r_N=1}^{i_N}\prod_{k=1}^NS_{i_k}^{(r_k)}(g_k)_{r_k}\bin{g_k-\delta_{k, \alpha}-r_k}{p_k-r_k}X_k^{p_k}
\end{eqnarray}
and in the same way
\begin{eqnarray}\label{mn2}
\mathscr{M}_{n,2}^{\alpha,\beta}&=&g_{\alpha}g_{\beta}\sum_{|\vec{p}|=Q}\sum_{i=0}^n\bin{n}{i}\left(D_0-D_{\alpha}\right)^{n-i}\sum_{i_1+i_2+\cdots i_N=i}\bin{i}{i_1,i_2,\cdots,i_N}\nonumber\\
& &\times\sum_{r_1=1}^{i_1}\sum_{r_2=1}^{i_2}\cdots\sum_{r_N=1}^{i_N}\prod_{k=1}^NS_{i_k}^{(r_k)}(g_k)_k\bin{g_k-\delta_{k,\alpha}-\delta_{k,\beta}-r_k}{p_k-r_k}X_k^{p_k}.
\end{eqnarray}
Such formulas do not contain any alternating summations. Their computation time is conditioned by the enumeration of partitions, for which efficient algorithms exist (see Appendix \ref{appB}). 

It is worth mentioning that expressions (\ref{mn1}) and (\ref{mn2}) can be simplified in the statistical-weight approximation (see Appendix \ref{appC}). This can be useful if one is interested in performing fast calculations at high temperature, for instance in the simulation of laser experiments in Hohlraums.

\subsection{Case of the second-order moment neglecting UTA widths}\label{subsec42}

The unnormalized, uncentered second-order moment reads, still neglecting the UTA width:
\begin{equation}
\mathscr{M}_{2}^{\alpha\beta}=\sum_{|\vec{p}|=Q}p_{\alpha}\left(g_{\beta}-p_{\beta}\right)\left[D_0+\sum_{s=1}^N\left(p_s-\delta_{s, \alpha}\right)D_s\right]^2\prod_{k=1}^N\bin{g_k}{p_k}X_k^{p_k}.
\end{equation}
Thanks to Eq. (\ref{identit}), one has
\begin{equation}
p_{\alpha}\left(g_{\beta}-p_{\beta}\right)=g_{\alpha}g_{\beta}\frac{\bin{g_{\beta}-1}{p_{\beta}}}{\bin{g_{\beta}}{p_{\beta}}}-g_{\alpha}g_{\beta}\frac{\bin{g_{\alpha}-1}{p_{\alpha}}\bin{g_{\beta}-1}{p_{\beta}}}{\bin{g_{\alpha}}{p_{\alpha}}\bin{g_{\beta}}{p_{\beta}}},
\end{equation}
and $\mathscr{M}_{2}^{\alpha\beta}$ can be put in the form
\begin{equation}
\mathscr{M}_2^{\alpha\beta}=\mathscr{M}_{2,1}^{\alpha\beta}-\mathscr{M}_{2,2}^{\alpha\beta}
\end{equation}
with
\begin{equation}
\mathscr{M}_{2,1}^{\alpha\beta}=g_{\alpha}g_{\beta}\sum_{|\vec{p}|=Q}\left[D_0+\sum_{s=1}^N\left(p_s-\delta_{s, \alpha}\right)D_s\right]^2\prod_{k=1}^N\bin{g_k-\delta_{k, \beta}}{p_k}X_k^{p_k} 
\end{equation}
and
\begin{equation}
\mathscr{M}_{2,2}^{\alpha\beta}=g_{\alpha}g_{\beta}\sum_{|\vec{p}|=Q}\left[D_0+\sum_{s=1}^N\left(p_s-\delta_{s, \alpha}\right)D_s\right]^2\prod_{k=1}^N\bin{g_k-\delta_{k, \alpha}-\delta_{k, \beta}}{p_k}X_k^{p_k}. 
\end{equation}
We have also
\begin{equation}
\left[D_0+\sum_{s=1}^N\left(p_s-\delta_{s, \alpha}\right)D_s\right]^2=\left(D_0-D_{\alpha}\right)^2+\left(\sum_{s=1}^Np_sD_s\right)^2+2\left(D_0-D_{\alpha}\right)\left(\sum_{s=1}^Np_sD_s\right)
\end{equation}
and
\begin{equation}
\left(\sum_{s=1}^Np_sD_s\right)^2=\sum_{s=1}^Np_s^2D_s^2+\sum_{\substack{s,s'=1\\s\ne s'}}^Np_sp_{s'}D_sD_{s'}.
\end{equation}
Using Eq. (\ref{stir}), we get that the final expression of the contribution $\mathscr{M}_{2,1}^{\alpha\beta}$ to the moment is therefore

\begin{eqnarray}
\mathscr{M}_{2,1}^{\alpha\beta}&=&g_{\alpha}g_{\beta}\sum_{|\vec{p}|=Q}\left\{\left(D_0-D_{\alpha}\right)^2+2\left(D_0-D_{\alpha}\right)\sum_{s=1}^Np_sD_s+\sum_{s=1}^Np_s^2D_s^2+\sum_{s,s'=1}^Np_sp_{s'}D_sD_{s'}\right\}\nonumber\\
& &\prod_{k=1}^N\bin{g_k-\delta_{k, \beta}}{p_k}X_k^{p_k},
\end{eqnarray}
yielding
\begin{eqnarray}
\mathscr{M}_{2,1}^{\alpha\beta}&=&g_{\alpha}g_{\beta}\sum_{|\vec{p}|=Q}\left(D_0-D_{\alpha}\right)^2\prod_{k=1}^N\bin{g_k-\delta_{k, \beta}}{p_k}X_k^{p_k}\nonumber\\
& &+2g_{\alpha}g_{\beta}\left(D_0-D_{\alpha}\right)\sum_{s=1}^ND_sS_1^{(1)}(g_s)_1\prod_{k=1}^N\bin{g_k-\delta_{k, \beta}-\delta_{k, s}}{p_k-\delta_{k, s}}X_k^{p_k}\nonumber\\
& &+g_{\alpha}g_{\beta}\sum_{s=1}^ND_s^2\sum_{r=1}^2S_2^{(r)}(g_s)_r\prod_{k=1}^N\bin{g_k-\delta_{k, \beta}-r\delta_{k, s}}{p_k-r\delta_{k, s}}X_k^{p_k}\nonumber\\
& &+g_{\alpha}g_{\beta}\sum_{\substack{s,s'=1\\s\ne s'}}^ND_sD_{s'}S_1^{(1)}\left(g_s\right)_1S_1^{(1)}\left(g_{s'}\right)_1\prod_{k=1}^N\bin{g_k-\delta_{k, \beta}-\delta_{k, s}-\delta_{k, s'}}{p_k-\delta_{k, s}-\delta_{k, s'}}X_k^{p_k}
\end{eqnarray}
or
\begin{eqnarray}
\mathscr{M}_{2,1}^{\alpha\beta}&=&g_{\alpha}g_{\beta}X_{\beta}\left(D_0-D_{\alpha}\right)^2\mathcal{U}_Q\left(g^{\beta}\right)\nonumber\\
& &+2g_{\alpha}g_{\beta}X_{\beta}\left(D_0-D_{\alpha}\right)\sum_{s=1}^NX_sD_s\mathcal{U}_{Q-1}\left(g^{\beta s}\right)\nonumber\\
& &+g_{\alpha}g_{\beta}\sum_{s=1}^ND_s^2\left[g_sX_s\mathcal{U}_{Q-1}\left(g^{\beta s}\right)+g_s\left(g_s-1\right)X_s^2\mathcal{U}_{Q-2}\left(g^{\beta s s}\right)\right]\nonumber\\
& &+g_{\alpha}g_{\beta}\sum_{\substack{s,s'=1\\s\ne s'}}^ND_sD_{s'}g_sg_{s'}X_sX_{s'}\mathcal{U}_{Q-2}\left(g^{\beta s s'}\right),
\end{eqnarray}
and in the same way
\begin{eqnarray}
\mathscr{M}_{2,2}^{\alpha\beta}&=&g_{\alpha}g_{\beta}X_{\alpha}X_{\beta}\left(D_0-D_{\alpha}\right)^2\mathcal{U}_Q\left(g^{\alpha\beta}\right)\nonumber\\
& &+2g_{\alpha}g_{\beta}X_{\alpha}X_{\beta}\left(D_0-D_{\alpha}\right)\sum_{s=1}^Ng_sX_sD_s\mathcal{U}_{Q-1}\left(g^{\alpha\beta s}\right)\nonumber\\
& &+g_{\alpha}g_{\beta}X_{\alpha}X_{\beta}\sum_{s=1}^ND_s^2\left[g_sX_s\mathcal{U}_{Q-1}\left(g^{\alpha\beta s}\right)+g_s\left(g_s-1\right)X_s^2\mathcal{U}_{Q-2}\left(g^{\alpha\beta s s}\right)\right]\nonumber\\
& &+g_{\alpha}g_{\beta}X_{\alpha}X_{\beta}\sum_{\substack{s,s'=1\\s\ne s'}}^ND_sD_{s'}g_sg_{s'}X_sX_{s'}\mathcal{U}_{Q-2}\left(g^{\alpha\beta s s'}\right).
\end{eqnarray}
The computation of moments can also benefit from fast exponentiation (for the computation of all the $X_i=e^{-\beta(\epsilon_i-\mu)}$, $i=1, \cdots, N$, see Appendix \ref{appD}).

\section{Conclusion}\label{sec5}

Reduced partition functions are an important element in the calculation of moments in the Super-Transition-Arrays statistical approach for opacity calculations of hot plasmas. The zero$^{th}$-order moment is the intensity, the first-order moment the average energy and the second-order moment the variance. All these reduced partition functions can be expressed using nominal partition functions, \emph{i.e.}, with the full set of supershell degeneracies, but this leads to alternating sums, which can be a source of numerical instability. Krief has shown that, for the average energy, this approach does not lead to numerical instabilities when the energy dispersion of the subshells in the supershell is less than 5\,$k_BT$ \cite{Krief2015}. This enables one to save much computational time when using the Bar-Shalom recurrence relation. For the second-order moment, however, the situation is different: the presence of alternating sums may be prohibitive. The best solution is probably to use the doubly-recursive relation (over the numbers of electrons and subshells), which applies to the partition functions with reduced (also called shifted or modified) degeneracies. In that case the numerical cost is higher, but the formulas remain simpler, and the number of alternating sums is smaller.

In the present work, we reviewed the different methods for the computation of partition functions with shifted degeneracies. We discussed the different precautions that must be taken, in order to avoid indeterminate forms. We also applied our recently-published expansion of partition functions (which accuracy can be controlled by truncating the series) to the reduced degeneracies, and obtained compact expressions. Finally, we provided positive-definite expressions of the STA moments of any order, without taking into the UTA width in a first attempt, based on the use of multinomial coefficients and Stirling numbers of the second kind. Such formulas, however, may have a high numerical cost, but can be optimized using efficient partitioning algorithms. The problem with computing STA machinery boils down to two possibilities:
\begin{itemize}
    \item (i) either one can use recursion relations, in which case alternating signs crop up, which lead to numerical instabilities and/or loss of accuracy, 
    \item (ii) or one can compute with formula that has only positive definite terms, but then one usually ends up with so many terms it is either too time consuming and/or memory intensive.
\end{itemize}
For example, from the original paper \cite{Barshalom1989} an example of the latter option was Eqs. (34) and (36), which required computations of partition functions of different sets of degeneracies, and an example of the former strategy was Eqs. (50) to (54), which necessitated alternating signs in Eq. (51). The Stirling numbers' approach exemplifies aspect (ii) and the double recurrence relation exemplifies aspect (i). The double recurrence relation for calculating partition functions by adding a shell at a time is a hybrid approach which avoids alternating signs but does not require too many operations. It would be nice to have a formalism that exhibits the same hybrid philosophy for moments. 

The calculation of partition function is inseparable from the construction of the supershells. For that purpose, enumerative combinatorics \cite{Cartier1969,Viennot1986} plays also a major role, since the corresponding supershell-generation algorithms often rely on successive splittings and gatherings of supershells, which usually involves different classes of binary trees.

Finally, in the future we plan to include effective statistical weights in the formalism, in order to obtain a smooth disappearance of the orbitals \cite{Busquet2013}, when pressure ionization takes place, in high-density plasmas.

\section*{Acknowledgements}

J.-C. Pain is indebted to Joseph Oreg for useful discussions at the 13$^{th}$ ``Radiative Properties of Hot Dense Matter'' conference in Santa Barbara, 10-14 November 2008. 

This work was performed under the auspices of the U.S. Department of Energy by Lawrence Livermore National Laboratory under Contract DE-AC52-07NA27344.


\section{Appendix A: Double recursion relation with decreasing number of electrons}\label{appA}

The generating-function technique used for the derivation of the double recursion relation (\ref{gilpain}) can be used in a ``reverse order'', \emph{i.e.}, expressing the partition function with $Q$ electrons and $N-1$ subshell in terms of the partition function with $Q-k$ electrons in $N$ subshell. The generating function reads
\begin{equation}
\delta_{0,y}=\frac{1}{2i\pi}\int_{-i\pi+\alpha_0}^{i\pi+\alpha_0}e^{ty}dt,
\end{equation}
where $y=Q-\sum_{s=1}^Np_s$ and $\alpha_0$ a real parameter. The partition function thus reads
\begin{equation}
\mathcal{U}_Q=\frac{1}{2i\pi}\int_{-i\pi+\alpha_0}^{i\pi+\alpha_0}\left[\prod_{s=1}^N\sum_{p_s=0}^{g_s}\bin{g_s}{p_s}\left(X_se^{-t}\right)^{p_s}\right]e^{tQ}dt.
\end{equation}
Setting $z=e^{-t}$, the integration in the complex plane is performed around a closed circle of radius $e^ {-\alpha_0}$, surrounding the pole at $z=0$. The latter expression becomes
\begin{equation}\label{diset}
\mathcal{U}_Q=\frac{1}{2i\pi}\oint\frac{\mathscr{F}_N(z)}{z^{Q+1}}dz,
\end{equation}
where $\mathscr{F}_N(z)$ is defined by
\begin{equation}
\mathscr{F}_N(z)=\prod_{s=1}^N\sum_{p_s=0}^{g_s}\bin{g_s}{p_s}\left(X_sz\right)^{p_s}=\prod_{s=1}^N\left(1+zX_s\right)^{g_s}.
\end{equation}
Equation (\ref{diset}) is calculated using the Cauchy formula, by evaluating the residue of the function $\mathscr{F}_N(z)/z^{Q+1}$ at the pole $z=0$ of order $Q+1$, \emph{i.e.}
\begin{equation}
\mathcal{U}_Q=\mathrm{Res}\left(\frac{\mathscr{F}_N(z)}{z^{Q+1}};0\right)=\lim_{z\rightarrow 0}\frac{1}{Q!}\frac{\partial^Q}{\partial z^Q}\mathscr{F}_N(z).
\end{equation}
One has therefore
\begin{equation}
\mathcal{U}_{Q,N-1}=\left.\frac{1}{Q!}\frac{\partial^Q}{\partial z^Q}\mathscr{F}_{N-1}(z)\right|_{z=0}=\left.\frac{1}{Q!}\frac{\partial^Q}{\partial z^Q}\left[(1+zX_N)^{-g_N}\prod_{s=1}^N\left(1+zX_s\right)^{g_s}\right]\right|_{z=0}.
\end{equation}
The latter result can also be obtained using the Egorychev method \cite{Egorychev,Wilf}, consisting in expressing the binomial coefficient as
\begin{equation}
\binom{n}{k}=\mathrm {Res}\left[\frac {(1+z)^{n}}{z^{k+1}};0\right]={\frac {1}{2\pi i}}\int _{|z|=\rho }{\frac {(1+z)^{n}}{z^{k+1}}}\;dz,
\end{equation}
where $0<\rho<\infty$. Applying the Leibniz formula for the derivative of a product, one gets
\begin{equation}
\mathcal{U}_{Q,N-1}=\left.\frac{1}{Q!}\sum_{k=0}^Q\bin{Q}{k}\frac{\partial^k}{\partial z^k}(1+zX_N)^{-g_N}.\frac{\partial^{Q-k}}{\partial z^{Q-k}}\prod_{s=1}^N\left(1+zX_s\right)^{g_s}\right].
\end{equation}
Calculating the $k^{th}$ derivative of $(1+zX_N)^{-g_N}$, one obtains
\begin{eqnarray}
\mathcal{U}_{Q,N-1}&=&\left.\frac{1}{Q!}\sum_{k=0}^Q\bin{Q}{k}(-1)^kg_N(g_N+1)(g_N+2)\cdots(g_N+k-1)X_N^k(1+zX_N)^{-g_N-k}\right.\nonumber\\
& &\left.\times\frac{\partial^{Q-k}}{\partial z^{Q-k}}\prod_{s=1}^N\left(1+zX_s\right)^{g_s}\right|_{z=0}.
\end{eqnarray}
The latter expression simplifies into
\begin{equation}
\mathcal{U}_{Q,N-1}=\left.\frac{1}{Q!}\sum_{k=0}^Q\frac{Q!}{k!(Q-k)!}\frac{(g_N+k-1)!}{(g_N-1)!}(-X_N)^k(1+zX_N)^{-g_N-k}.\frac{\partial^{Q-k}}{\partial z^{Q-k}}\prod_{s=1}^N\left(1+zX_s\right)^{g_s}\right|_{z=0},
\end{equation}
leading to
\begin{equation}
\mathcal{U}_{Q,N-1}=\left.\sum_{k=0}^Q\frac{(g_N+k-1)!}{k!(g_N-1)!}(-X_N)^k(1+zX_N)^{-g_N-k}\frac{1}{(Q-k)!}\frac{\partial^{Q-k}}{\partial z^{Q-k}}\prod_{s=1}^N\left(1+zX_s\right)^{g_s}\right|_{z=0},
\end{equation}
yielding
\begin{equation}
\mathcal{U}_{Q,N-1}=\sum_{k=0}^Q\bin{g_N+k-1}{g_N-1}(-X_N)^k\:\mathcal{U}_{Q-k,N-1},
\end{equation}
involving, as Oreg's relation (\ref{oreg}), the number of combinations with repetitions of $k$ elements among $g_N$. Unfortunately, the relation is not very efficient since it involve terms with alternate signs. This happens rather often. For instance, relation (\ref{oreg}) contains alternate-sign terms, but the relation expressing $\mathcal{U}_Q(g)$ in terms of partition functions of the kind $\mathcal{U}_Q\left(g^a\right)$ does not \cite{Oreg1997}, as illustrated, for instance, by the relations:
\begin{equation}
\mathcal{U}_Q(g)=\frac{1}{G-Q}\sum_{s=1}^Ng_s\mathcal{U}_Q\left(g^s\right)
\end{equation}
as well as
\begin{equation}
\mathcal{U}_Q(g)=\frac{1}{Q}\sum_{s=1}^Ng_sX_s\mathcal{U}_{Q-1}\left(g^s\right)
\end{equation}
and
\begin{equation}
\mathcal{U}_{Q}(g)=\frac{1}{Q^2}\sum_{r,s=1}^Ng_r\left(g_s-\delta_{r,s}\right)X_rX_s\mathcal{U}_{Q-2}\left(g^{rs}\right).
\end{equation}

\section{Appendix B: Partitions and multinomial coefficients}\label{appB}

Let $q$ and $n$ be two integers, $q\geq 1$, and $k_1, k_2,..., k_q$ are real numbers. Then,
\begin{equation}
\left(x_{1}+x_{2}+x_{3}+\dots +x_{q}\right)^{n}=\sum _{k_{1}+k_{2}+k_{3}+\ldots +k_{q}=n}\binom{n}{k_{1},k_{2},k_{3},\dots ,k_{q}}x_{1}^{k_{1}}x_{2}^{k_{2}}x_{3}^{k_{3}}\dots x_{q}^{k_{q}}.
\end{equation}
The sum covers all $q$ tuples of natural numbers $(k_1, k_2,...,k_q)$ such that $k_1 + k_2 + ... + k_q = n$, some of which may be zero. If we arrange the multinomial coefficients in a triangle so that in row $n$ are the 
\begin{equation}
\binom{n}{k_{1},k_{2},\ldots ,k_{q}},
\end{equation}
with $k_{1}\geq k_{2}\geq \dots \geq k_{q}\geq 1$, the $\left(k_{1},k_{2},\dots ,k_{q}\right)$ are arranged in descending lexicographical order, we obtain the first lines, starting with $n =1$:

\begin{table}[!ht]
\begin{tabular}{l}
1, 2\\
1, 3, 6\\
1, 4, 6, 12, 24\\
1, 5, 10, 20, 30, 60, 120\\
1, 6, 15, 20, 30, 60, 90, 120, 180, 360, 720
\end{tabular}
\caption{\label{tab:table3} first values of the multinomial coefficient sorted by rows $n$ (first line corresponds to $n=$1, second line to $n=$2, \emph{etc.}).} 
\end{table}

Note that in this triangle, the number of terms in line $n$ is equal to the number $p(n)$ of partitions of the integer $n$; the sum of the terms of a line is listed as OEIS sequence A005651 \cite{OEIS}.

The total number of terms in the expansion of $\left(\sum _{i=1}^{q}x_{i}\right)^{n}$ is equal to the number of unitary monomials of degree $n$ formed from $x_1$, $x_2$,..., $x_q$, i.e. the number of their $n$-combinations with repetitions 
\begin{equation}
\binom{n+q-1}{q-1}.
\end{equation}
Denoting, for $n$ and $q$ strictly positive integers, $p(n,q)$ the number of partitions of $n$ in $q$ parts, the function $p$ is recursive and verifies the following relationship for all $n > q > 1$:
\begin{equation}
p(n, q) = p(n - 1, q - 1) + p(n - q, q),
\end{equation}
with the initial conditions $p(n, q) = 0$ if $n < q$, $p(n, n) = p(n, 1) = 1$. The relationship arises from a distinction of cases among these partitions: either the last (smallest) part is worth 1, in which case the partition is obtained from a partition of $(n - 1)$ in $(q - 1)$ parts, by adding this last part; or all parts are worth at least 2, in which case the partition is obtained from a partition of $(n - q)$ into $q$ parts, by increasing each part by one.

Provided that ``memoization'' is used, this procedure enables the number of partitions of an integer to be calculated with quadratic algorithmic complexity as a function of $n$, by adding up all the values of $p(n, q)$ when $q$ varies between 1 and $n$. In programming, memoization is an optimization technique that makes applications more efficient and hence faster \cite{Michie1968}. It does this by storing computation results in cache, and retrieving that same information from the cache the next time it's needed instead of computing it again. In simpler words, it consists of storing in cache the output of a function, and making the function check if each required computation is in the cache before computing it. A cache is simply a temporary data store that holds data so that future requests for that data can be served faster. Memoization is a simple but powerful trick that can help speed up our code, especially when dealing with repetitive and heavy computing function. Consider the following function calculating the terms of the Lucas sequence \cite{Borwein1987}:
\begin{verbatim}
luc(n) {
   if n is equal to 0 then
       return 2
   else
     if n is equal to 1 then
         return 1
     else
         return luc(n-1) + luc(n-2);
     endif
   endif
} 
\end{verbatim}
Lucas number $L_n$ is equal to $\phi^n+(1-\phi)^n$, where $\phi$ is the golden ratio. As it stands, this recursive function is extremely inefficient (time complexity O($\phi^{n}$) where $\phi$ is the golden ratio), as many recursive calls are made on the same values of $n$. The ``memoized'' version of \verb!luc! stores previously calculated values in an associative table:

\begin{verbatim}
memoluc(n) {
   if table(n) undefined then
     if n is equal to 0 then
         table(n)=2
     else
       if n is equal to 1 then
           table(n)=1
       else
           table(n) = luc(n-1) + luc(n-2);
       endif
     endif
   return table(n);
}
\end{verbatim}

The function calculates the value \verb!table(n)! if it has not yet been defined, then returns the value stored in \verb!table(n)!. The complexity of \verb!memoluc! is linear in both time and space. Note that there are even more efficient ways of calculating the terms of the Fibonacci sequence, but this is just to illustrate \verb!memoluc!.

A more efficient way of calculating the number of partitions of an integer is deduced from Euler's pentagonal number theorem. This gives a recurrence relation which can be written as
\begin{equation}
p(n)=p(n-1)+p(n-2)-p(n-5)-p(n-7)+p(n-12)+p(n-15)+\dots =\sum _{q\geq 1}(-1)^{q-1}p\left(n-\frac{q(3q\pm 1)}{2}\right).
\end{equation}
The $q(3q\pm 1)/2$ are generalized pentagonal numbers. For instance, the number of multinomial coefficients 
\begin{equation}
\binom{4}{k_1,k_2,k_3}
\end{equation}
in a sum $(a+b+c)^4$ is equal to
\begin{equation}
\binom{4+3-1}{3-1}=\binom{6}{2}=15,
\end{equation}
corresponding to $(4,0,0)$, $(0,4,0)$, $(0,0,4)$, $(3, 1, 0)$, $(3,0,1)$, $(1,3,0)$, $(0,3,1)$, $(1,0,3)$, $(0,1,3)$, $(2,2,0)$, $(0,2,0)$, $(0,0,2)$, $(2,1,1)$, $(1,2,1)$ and $(1,1,2)$, while the number of partition of 4 into 3 sets, i.e. $p(4,3)$ is equal to 1, and corresponds to $(1,1,2)$.

\section{Appendix C: Moments in the statistical-weight approximation}\label{appC}

In the framework of the design or interpretation of inertial-confinement-fusion experiments, inline calculations of opacity and emissivity are required at each time step and in each spatial cell, with the corresponding radiation field, of radiative-hydrodynamics simulations of Hohlraums. The computation time is therefore the main limiting factor, and developing fast approximate methods for the calculation of partition functions would enable one to investigate the possibility of carrying out, in such complex simulations, the collisional-radiative modeling of the plasma in the superconfiguration approximation \cite{Barshalom2000}. Indeed, the computation of the rates of the different radiative and collisional involved processes requires the determination of the canonical partition functions of the supershells. At high temperature, it can be relevant to simplify the partition function by replacing the $X_i$ factors by the degeneracy $g_i$. In that case, the partition functions reduce to simple binomial coefficients.

The final expression of the contribution $\mathscr{M}_{n,1}^{\alpha\beta}$ to the moment becomes then

\begin{eqnarray}
\mathscr{M}_{n,1}^{\alpha\beta}&=&g_{\alpha}g_{\beta}\sum_{|\vec{p}|=Q}\sum_{i=0}^n\bin{n}{i}\left(D_0-D_{\alpha}\right)^{n-i}\sum_{i_1+i_2+\cdots+i_N=i}\bin{i}{i_1,i_2,\cdots,i_N}\nonumber\\
& &\times D_1^{i_1}D_2^{i_2}\cdots D_N^{i_N}\sum_{r_1=1}^{i_1}\sum_{r_2=1}^{i_2}\cdots\sum_{r_N=1}^{i_N}\prod_{k=1}^NS_{i_k}^{(r_k)}(g_k)_{r_k}\bin{g_k-\delta_{k, \alpha}-r_k}{p_k-r_k}X_k^{p_k},
\end{eqnarray}
\textit{i.e.}, in the idealized statistical-weight approximation (infinite temperature, $\beta=0$ and thus, $\forall k$, $X_k=1$):
\begin{eqnarray}
\mathscr{M}_{n,1}^{\alpha,\beta}&=&g_{\alpha}g_{\beta}\sum_{i=0}^n\bin{n}{i}\left(D_0-D_{\alpha}\right)^{n-i}\sum_{i_1+i_2+\cdots i_N=i}\bin{i}{i_1,i_2,\cdots,i_N}\nonumber\\
& &\times\sum_{r_1=1}^{i_1}\sum_{r_2=1}^{i_2}\cdots\sum_{r_N=1}^{i_N}\prod_{k=1}^NS_{i_k}^{(r_k)}(g_k)_k\bin{G-1-\sum_{k=1}^Nr_k}{Q-\sum_{k=1}^Nr_k},
\end{eqnarray}
and in the same way
\begin{eqnarray}
\mathscr{M}_{n,2}^{\alpha,\beta}&=&g_{\alpha}g_{\beta}\sum_{|\vec{p}|=Q}\sum_{i=0}^n\bin{n}{i}\left(D_0-D_{\alpha}\right)^{n-i}\sum_{i_1+i_2+\cdots i_N=i}\bin{i}{i_1,i_2,\cdots,i_N}\nonumber\\
& &\times\sum_{r_1=1}^{i_1}\sum_{r_2=1}^{i_2}\cdots\sum_{r_N=1}^{i_N}\prod_{k=1}^NS_{i_k}^{(r_k)}(g_k)_k\bin{g_k-\delta_{k,\alpha}-\delta_{k,\beta}-r_k}{p_k-r_k}X_k^{p_k},
\end{eqnarray}
giving, in the statistical-weight approximation ($\forall k$, $X_k=1$):
\begin{eqnarray}
\mathscr{M}_{n,2}^{\alpha,\beta}&=&g_{\alpha}g_{\beta}\sum_{i=0}^n\bin{n}{i}\left(D_0-D_{\alpha}\right)^{n-i}\sum_{i_1+i_2+\cdots i_N=i}\bin{i}{i_1,i_2,\cdots,i_N}\nonumber\\
& &\times\sum_{r_1=1}^{i_1}\sum_{r_2=1}^{i_2}\cdots\sum_{r_N=1}^{i_N}\prod_{k=1}^NS_{i_k}^{(r_k)}(g_k)_k\bin{G-2-\sum_{k=1}^Nr_k}{Q-\sum_{k=1}^Nr_k}
\end{eqnarray}
and since
\begin{equation}
\bin{n}{p}=\bin{n-1}{p}+\bin{n-1}{p-1},
\end{equation}
we have
\begin{eqnarray}
\mathscr{M}_{n}^{\alpha,\beta}&=&g_{\alpha}g_{\beta}\sum_{i=0}^n\bin{n}{i}\left(D_0-D_{\alpha}\right)^{n-i}\sum_{i_1+i_2+\cdots i_N=i}\bin{i}{i_1,i_2,\cdots,i_N}\nonumber\\
& &\times\sum_{r_1=1}^{i_1}\sum_{r_2=1}^{i_2}\cdots\sum_{r_N=1}^{i_N}\prod_{k=1}^NS_{i_k}^{(r_k)}(g_k)_k\bin{G-1-\sum_{k=1}^Nr_k}{Q-1-\sum_{k=1}^Nr_k}.
\end{eqnarray}

\section{Appendix D: Speeding up the calculation}\label{appD}

\subsection{Fast exponentiation and polynomial multiplication}

The partition functions involve many powers of $X_i$, the exponent being the population (number of electrons) of the subshell $i$. It is useful, in order to hasten the computations, to resort to fast exponentiation (or ``square-and-multiply'' or Chandah-sutra \cite{OEISA014701}) techniques. The first way of obtaining $n^p$ ($n$ and $p$ being integers) is of course to multiply $n$ by itself $p$ times. However, there are far more efficient methods, where the number of operations required is no longer of the order of $p$, but of the order of $\ln(p)$. For instance, if we write
\begin{equation}
p=\sum_{i\leq d}a_{i}2^{i}
\end{equation}
for $a_{i}\in \{0,1\}$, we notice that
\begin{equation}
n^{p}=n^{a_{0}}\left(n^{2}\right)^{a_{1}}\left(n^{2^{2}}\right)^{a_{2}}\dots \left(n^{2^{d}}\right)^{a_{d}}. 
\end{equation}
It thus takes $d$ operations to calculate all $n^{2^{i}}$, then $d$ additional operations to form the product of $\left(n^{2^{i}}\right)^{a_{i}}$. The total number of operations is therefore $2d$, which is of the order of the logarithm of $p$. This simple algebraic remark leads to the following algorithm. Let $n$ be an integer strictly greater than 1. Suppose we know how to calculate, for each real $x$, all the powers $x^k$ of $x$, for all $k$, such that $1 \leq k < n$. If $n$ is even, then $x^n = (x^2)^{n/2}$. It is then sufficient to calculate $y^{n/2}$ for $y = x^2$. If $n$ is odd and $n > 1$, then $x^n = x(x^2)^{(n - 1)/2}$, and one can simply calculate $y^{(n - 1)/2}$ for $y = x^2$ and multiply by $x$. This leads us to the following recursive algorithm, which calculates $x^n$ for a strictly positive integer $n$:
\begin{equation}
{\mbox{power}}(x,\,n)=\left\{{\begin{matrix}x,&{\mbox{if }}n{\mbox{ = 1}}\\{\mbox{power}}\left(x^{2},\,\frac{n}{2}\right),&{\mbox{if }}n{\mbox{ is even}}\\x\times {\mbox{power}}\left(x^{2},\,\frac{(n-1)}{2}\right),&{\mbox{if }}n{\mbox{ is odd}}.\\\end{matrix}}\right.
\end{equation}
Compared with the ordinary method of multiplying $x$ by itself $n - 1$ times, this algorithm requires $O(\log n)$ multiplications, which speeds up the calculation of $x^n$ dramatically for large integers. The algorithm is implemented in the short {\sc Fortran} 90 program below.
\begin{verbatim}
module fast_exp
implicit none

interface operator (.fexp.) 
  module procedure realexp
end interface

contains

  function realexp (base, exponent)
    real :: realexp
    real, intent(in) :: base
    integer, intent(in) :: exponent
    integer :: i
  
    realexp = 1.0
    if (exponent < 0) then
       do i = exponent, -1
          realexp = realexp / base
       end do
    else  
       do i = 1, exponent
          realexp = realexp * base
       end do
    end if
  end function realexp

end module fast_exp

program example
use fast_exp
  write(*,*) 3.4.fexp.10
end program example
\end{verbatim}
Using ** instead of ``\verb!.fexp.!'' on the third line would overload the standard exponentiation operator. The output of the above program is 206437.812.

Since the partition functions can be viewed as multivariate polynomials (the variables being $X_i$, $i=1, \cdots, N$), the computations can also be sped up using well-known techniques, such as the Karatsuba algorithm for products of polynomials \cite{Karatsuba1963}, or other divide-and-conquer algorithms for euclidean division for instance \cite{Knuth1969,Hart2011}.

\subsection{Matrix representation}

The relation (\ref{oreg}) between partition functions can be put in a matrix form, which can be expressed as
\begin{equation}
\mathcal{U}_Q(\bar{g}^a)=\sum_{k=0}^Q{\bf N}_{Q,k}~\mathcal{U}_k(g),
\end{equation}
with 
\begin{equation}
{\bf N}_{Q,k}=N_{k,Q}^{(g_a)}(-X_a)^{Q-k},
\end{equation}
where we introduce the dependence with respect to $Q$, which was hidden in $N_k^{(m)}$ as noted by Oreg (see Eq. (\ref{oregnot})). For instance, in a subshell with degeneracy $G=6$, one has
\begin{equation}
\left(
\begin{array}{c}
\mathcal{U}_0(\bar{g}^a)\\
\mathcal{U}_1(\bar{g}^a)\\
\mathcal{U}_2(\bar{g}^a)\\
\mathcal{U}_3(\bar{g}^a)\\
\mathcal{U}_4(\bar{g}^a)\\
\mathcal{U}_5(\bar{g}^a)\\
\mathcal{U}_6(\bar{g}^a)\\
\end{array}
\right)=\left[
\begin{array}{ccccccc}
1 & 0 & 0 & 0 & 0 & 0 & 0\\
{\bf N}_{1,0} & {\bf N}_{1,1} & 0 & 0 & 0 & 0 & 0\\
{\bf N}_{2,0} & {\bf N}_{2,1} & {\bf N}_{2,2} & 0 & 0 & 0 & 0\\
{\bf N}_{3,0} & {\bf N}_{3,1} & {\bf N}_{3,2} & {\bf N}_{3,3} & 0 & 0 & 0\\
{\bf N}_{4,0} & {\bf N}_{4,1} & {\bf N}_{4,2} & {\bf N}_{4,3} & {\bf N}_{4,4} & 0 & 0\\
{\bf N}_{5,0} & {\bf N}_{5,1} & {\bf N}_{5,2} & {\bf N}_{5,3} & {\bf N}_{5,4} & {\bf N}_{5,5} & 0\\
{\bf N}_{6,0} & {\bf N}_{6,1} & {\bf N}_{6,2} & {\bf N}_{6,3} & {\bf N}_{6,4} & {\bf N}_{6,5} & {\bf N}_{6,6}\\
\end{array}
\right]
\left(
\begin{array}{c}
\mathcal{U}_0(g)\\
\mathcal{U}_1(g)\\
\mathcal{U}_2(g)\\
\mathcal{U}_3(g)\\
\mathcal{U}_4(g)\\
\mathcal{U}_5(g)\\
\mathcal{U}_6(g)\\
\end{array}
\right),
\end{equation}
which enables one to obtain $\mathcal{U}_Q\left(\bar{g}^{ab}\right)$, $\mathcal{U}_Q\left(\bar{g}^{abc}\right)$, \emph{etc.} by successive multiplications of matrices. Note that the above matrix relation can be inverted as
\begin{equation}
\left(
\begin{array}{c}
\mathcal{U}_0(g)\\
\mathcal{U}_1(g)\\
\mathcal{U}_2(g)\\
\mathcal{U}_3(g)\\
\mathcal{U}_4(g)\\
\mathcal{U}_5(g)\\
\mathcal{U}_6(g)\\
\end{array}
\right)=\left[
\begin{array}{ccccccc}
1 & 0 & 0 & 0 & 0 & 0 & 0\\
{\bf Y}_{1,0} & 1 & 0 & 0 & 0 & 0 & 0\\
{\bf Y}_{2,0} & {\bf Y}_{2,1} & 1 & 0 & 0 & 0 & 0\\
{\bf Y}_{3,0} & {\bf Y}_{3,1} & {\bf Y}_{3,2} & 1 & 0 & 0 & 0\\
{\bf Y}_{4,0} & {\bf Y}_{4,1} & {\bf Y}_{4,2} & {\bf Y}_{4,3} & 1 & 0 & 0\\
{\bf Y}_{5,0} & {\bf Y}_{5,1} & {\bf Y}_{5,2} & {\bf Y}_{5,3} & {\bf Y}_{5,4} & 1 & 0\\
{\bf Y}_{6,0} & {\bf Y}_{6,1} & {\bf Y}_{6,2} & {\bf Y}_{6,3} & {\bf Y}_{6,4} & {\bf Y}_{6,5} & 1\\
\end{array}
\right]
\left(
\begin{array}{c}
\mathcal{U}_0(\bar{g}^a)\\
\mathcal{U}_1(\bar{g}^a)\\
\mathcal{U}_2(\bar{g}^a)\\
\mathcal{U}_3(\bar{g}^a)\\
\mathcal{U}_4(\bar{g}^a)\\
\mathcal{U}_5(\bar{g}^a)\\
\mathcal{U}_6(\bar{g}^a)\\
\end{array}
\right),
\end{equation}
where
\begin{equation}
{\bf Y}_{Q,k}=\frac{X_a^{Q-k}}{(Q-k)!}g_a(g_a-1)\cdots (g_a-Q+k+1)=X_a^{Q-k}\binom{g_a}{Q-k}.
\end{equation}
and ${\bf Y}_{Q,Q}=1$, \emph{i.e.},
\begin{equation}
{\bf Y}_{Q,k}=X_a^{Q-k}\binom{g_a}{Q-k}.
\end{equation}
We thus have, by the way,
\begin{equation}
\mathcal{U}_Q(\bar{g}^a)=\sum_{k=0}^Q\bin{g_a}{Q-k}X_a^{Q-k}~\mathcal{U}_k(g).
\end{equation}
Strassen first suggested an algorithm \cite{Strassen1969} to multiply matrices with worst case running time less than the conventional $O(n^3)$ operations. He also presented a recursive algorithm dedicated to the inversion of matrices and the calculation of determinants using matrix multiplication. Bunch and Hopcroft improved the algorithm in 1974 \cite{Bunch1974} in the case where principal submatrices are singular. Recently, Tonks \emph{et al.} covered the case of multivariate polynomial matrix inversion \cite{Tonks2017}. The inverse of a matrix of polynomials is in general a matrix of rational functions. In performing these algorithms symbolically, this means we are forced to take greatest common divisors of potentially large polynomials in order to simplify results of inversion. The worst case complexity of taking the greatest common divisor is polynomial in the degree of the polynomials and exponential in the number of variables \cite{Brown1971}. As far as we are concerned the number of variables may as well be fixed, but the degree of numerators and denominators will increase per recursion as a result of the matrix arithmetic performed in any one recursion.

The automatic discovery of algorithms using machine learning offers the prospect of reaching beyond human intuition and outperforming the current best human-designed algorithms. However, automating the algorithm discovery procedure is intricate, as the space of possible algorithms is enormous. It is worth mentioning that Fawzi \emph{et al.} reported a deep reinforcement learning approach for discovering efficient algorithms for the multiplication of arbitrary matrices, based on training to find tensor decompositions within a finite factor space \cite{Fawzi2022}.

\end{document}